%% file: lyaauto.tex
\documentclass[traditabstract]{aa} 
\usepackage{graphicx}
\usepackage{txfonts}
\usepackage{natbib}
\usepackage{lineno}
\newcommand{\apar}{\alpha_\parallel}
\newcommand{\aperp}{\alpha_\perp}
\newcommand{\hMpc}{h^{-1}{\rm Mpc}}
\newcommand{\lya}{Ly$\alpha$}
\newcommand{\Lya}{Ly$\alpha$~}
\newcommand{\Lyb}{Ly$\beta$~}
\newcommand{\om}{\Omega_M}
\newcommand{\oc}{\Omega_C}
\newcommand{\ob}{\Omega_B}
\newcommand{\on}{\Omega_\nu}
\newcommand{\ol}{\Omega_\Lambda}

\newcommand{\rhode}{\rho_{\rm de}}
\newcommand{\dA}{D_A}
\newcommand{\dH}{D_H}
\newcommand{\dv}{D_V}

\newcommand{\rperp}{r_\perp}
\newcommand{\rpar}{r_\parallel}
\newcommand{\lcdm}{$\Lambda$CDM~}
\newcommand{\hubunits}{\,{\rm km}\,{\rm s}^{-1}\,{\rm Mpc}^{-1}}

\begin{document}
\newcommand{\asqq}[1]{[??AS: \textit{#1}]}
\newcommand{\jrqq}[1]{[??JR: \textit{#1}]}
\newcommand{\mbqq}[1]{[??MB: \textit{#1}]}
\newcommand{\dwqq}[1]{[??DW: \textit{#1}]}
\newcommand{\mpqq}[1]{[??MP: \textit{#1}]}
   \title{Baryon acoustic oscillations in the \Lya forest of BOSS DR11 quasars}

\author{
Timoth\'ee~Delubac\inst{1,12}, 
Julian~E.~Bautista\inst{2},
Nicol\'as~G.~Busca\inst{2,24,25}, 
James~Rich\inst{1},
David~Kirkby\inst{3},
Stephen~Bailey\inst{4},
Andreu~Font-Ribera\inst{4},
An\v{z}e~Slosar\inst{5},
Khee-Gan~Lee\inst{6},
Matthew~M.~Pieri\inst{7},
Jean-Christophe~Hamilton\inst{2},
\'{E}ric Aubourg\inst{2},
Michael~Blomqvist\inst{3},
Jo~Bovy\inst{8},
J. Brinkmann\inst{9},
William~Carithers\inst{4},
Kyle S. Dawson\inst{10},
Daniel~J.~Eisenstein\inst{11},
Satya~Gontcho A Gontcho\inst{15},
Jean-Paul Kneib\inst{12,13},
J.-M.~Le~Goff\inst{1}, 
Daniel~Margala\inst{3},
Jordi~Miralda-Escud\'{e}\inst{14,15},
Adam~D.~Myers\inst{16},
Robert~C.~Nichol\inst{7},
Pasquier~Noterdaeme\inst{17},
Ross~O'Connell\inst{18},
Matthew D. Olmstead\inst{10},
Nathalie~Palanque-Delabrouille\inst{1}, 
Isabelle~P\^aris\inst{17},
Patrick~Petitjean\inst{17},
Nicholas~P.~Ross\inst{4,19},
Graziano~Rossi\inst{1,26},
David~J.~Schlegel\inst{4},
Donald~P.~Schneider \inst{20,21},
David~H.~Weinberg\inst{22},
Christophe~Y\`eche\inst{1},
Donald G. York\inst{23} 
}
\institute{
CEA, Centre de Saclay, IRFU,  F-91191 Gif-sur-Yvette, France
\and
APC, Universit\'{e} Paris Diderot-Paris 7, CNRS/IN2P3, CEA,
Observatoire de Paris, 10, rue A. Domon \& L. Duquet,  Paris, France
\and
Department of Physics and Astronomy, University of California, Irvine, 
CA 92697, USA
\and
Lawrence Berkeley National Laboratory, 1 Cyclotron Road, Berkeley, CA 94720, USA
\and
Brookhaven National Laboratory, 2 Center Road,  Upton, NY 11973, USA
\and
Max-Planck-Institut f\"ur Astronomie, K\"onigstuhl 17, D69117 Heidelberg, Germany
\and
Institute of Cosmology and Gravitation, Dennis Sciama Building, University of Portsmouth, Portsmouth, PO1 3FX, UK
\and
 Institute for Advanced Study, Einstein Drive, Princeton, NJ 08540, USA
\and
Apache Point Observatory, P.O. Box 59, Sunspot, NM 88349, USA
\and
Department of Physics and Astronomy, University of Utah, 115 S 1400 E, Salt Lake City, UT 84112, USA
\and
Harvard-Smithsonian Center for Astrophysics, Harvard University, 60 Garden St., Cambridge MA 02138, USA
\and
Laboratoire d'Astrophysique, Ecole polytechnique Fed\'erale de Lausanne, CH-1015 Lausanne, Switzerland
\and
Aix Marseille Universit\'e, CNRS, LAM (Laboratoire d’Astrophysique de Marseille) UMR 7326, F-13388, Marseille, France
\and
Instituci\'{o} Catalana de Recerca i Estudis  Avan\c{c}ats, Barcelona, Catalonia
\and
Institut de Ci\`{e}ncies del Cosmos, Universitat de Barcelona (UB-IEEC), Catalonia
\and
Department of Physics and Astronomy, University of Wyoming, Laramie, WY 82071, USA
\and 
Universit\'e Paris 6 et CNRS, Institut d'Astrophysique de Paris, 98bis blvd. Ara
go, 75014 Paris, France
\and
Bruce and Astrid McWilliams Center for Cosmology, Carnegie Mellon University, Pittsburgh, PA 15213, USA
\and
Dept. of Physics, Drexel University, 3141 Chestnut Street, Philiadelphia, PA 19104, USA
\and
Department of Astronomy and Astrophysics, The Pennsylvania State University, University Park, PA 16802
\and
 Institute for Gravitation and the Cosmos, The Pennsylvania State University, University Park, PA 16802
\and
Department of Astronomy, Ohio State University, 140 West 18th Avenue, Columbus, 
OH 43210, USA
\and
Department of Astronomy and Astrophysics and the Enrico Fermi Institute, The University of Chicago, 5640 South Ellis Avenue, Chicago, Illinois, 60615, USA
\and
Observat\'orio Nacional, Rua Gal.~Jos\'e Cristino 77, Rio de Janeiro, RJ - 20921-400, Brazil
\and
Laborat\'orio Interinstitucional de e-Astronomia, - LIneA, Rua Gal.Jos\'e Cristino 77, Rio de Janeiro, RJ - 20921-400, Brazil  
\and
Department of Astronomy and Space Science, Sejong University, Seoul, 143-747, Korea
}

\date{Received April 9, 2014; accepted October 15, 2014}

\abstract{
We report a detection of the baryon acoustic oscillation (BAO)
feature in the flux-correlation function  
of the \Lya forest of high-redshift quasars with a statistical
significance of five standard deviations.
The study uses 137,562 quasars in the redshift range $2.1\le z \le 3.5$
from the Data Release 11 (DR11) of the
Baryon  Oscillation Spectroscopic Survey (BOSS) 
of SDSS-III.
This sample contains three times the number of quasars used
in previous studies.
The measured position of the BAO peak determines
the angular distance, $\dA(z=2.34)$ 
and expansion rate, $H(z=2.34)$, both on a scale
set by the sound horizon at the drag epoch, $r_d$.
We find $\dA/r_d=11.28\pm0.65(1\sigma)^{+2.8}_{-1.2}(2\sigma)$ and
$\dH/r_d=9.18\pm0.28(1\sigma)\pm0.6(2\sigma)$ where $\dH=c/H$.
The optimal combination, $\sim\dH^{0.7}\dA^{0.3}/r_d$ is determined with
a precision of $\sim2\%$.
For the value  $r_d=147.4~{\rm Mpc}$, 
consistent with the cosmic microwave background power spectrum
measured by Planck, 
we find
$\dA(z=2.34)=1662\pm96(1\sigma)~{\rm Mpc}$
 and 
$H(z=2.34)=222\pm7(1\sigma)~{\rm km\,s^{-1}Mpc^{-1}}$.
Tests with mock catalogs and variations of our analysis procedure
have revealed no systematic uncertainties comparable to our statistical
errors.
Our results agree
with the previously reported
BAO measurement at the same redshift using the quasar-\Lya
forest cross-correlation.
The autocorrelation and cross-correlation approaches are 
complementary because of the quite different impact of redshift-space
distortion on the two measurements.
The combined constraints from the two correlation functions imply
values of $\dA/r_d$ that are 7\% lower and 7\% higher for $\dH/r_d$ than the predictions of a flat \lcdm cosmological
model with the best-fit Planck parameters.
With our estimated statistical errors, the significance of this 
discrepancy is $\approx 2.5\sigma$.
}


\authorrunning{T. Delubac et al.}
\titlerunning{BAO in the \Lya forest of BOSS quasars}

\maketitle
%

\section{Introduction}
\label{introsec}

Observation of
the peak in the matter correlation function due to baryon acoustic oscillations
(BAO) in the pre-recombination epoch is now an established
tool to constrain cosmological models.
The BAO peak at a redshift $z$ appears at
an angular separation $\Delta\theta=r_d/[(1+z)\dA(z)]$ 
and at a redshift separation
$\Delta z=r_d/\dH(z)$, where $\dA$ and $\dH=c/H$ are the angular 
and Hubble distances, and $r_d$ is the sound horizon 
at the drag epoch\footnote{
We follow the convention of \citet{anderson13}, $r_d=r_s(z_d)$, where
$r_s$ is the sound horizon and
$z_d$ is the drag redshift (baryon decoupling from photons), to
be distinguished from $z_*$ (the redshift corresponding to unity optical
depth for CMB photons).  
Earlier
publications on BAO generally denoted $r_d$  simply as $r_s$.
For models with cold dark matter, baryons and three light neutrino
species, $r_d$ can be evaluated with
Eq. (55) of \citet{anderson13}, which agrees with the CAMB-derived
value to better than 0.1 per cent.
}.
Measurement of the peak position at any redshift thus
constrains the combinations of  cosmological parameters that
determine $\dH/r_d$ and $\dA/r_d$.

The BAO peak has been observed primarily in the galaxy-galaxy correlation
function obtained in redshift surveys.
The small statistical significance of the first studies gave only
constraints on $\dv/r_d$ where $\dv$ is the combination 
$\dv=[(1+z)\dA]^{2/3}[z\dH]^{1/3}$, which
determines the peak position for the galaxy correlation
function when averaged over directions with respect to the line of sight.
The first measurements were at $z\sim 0.3$ 
by the SDSS \citep{sdss1bao} and 2dFGRS \citep{2dfbao} with
results from the combined data set presented by \citet{percival10}.
A refined analysis using reconstruction
\citep{eisenstein07,padmanabhan09} to improve
the precision $\dv/r_d$ was presented by 
\citet{padmanabhan12} and \citet{mehta12}.

Other measurements of $\dv/r_d$ were made 
at $z\sim0.1$ by the 6dFGRS \citep{beutler11},
at  $(0.4<z<0.8)$ by WiggleZ \citep{blake11}, and,
using galaxy clusters, at $z\sim0.3$ by \citet{veropalumbo13}.
The Baryon Oscillation Spectroscopic Survey (BOSS;
\citealt{bossoverview}) of SDSS-III
\citep{eisenstein11}
has presented measurements  of $\dv/r_d$ 
at $z\sim 0.57$ and $z\sim0.32$ \citep{anderson12}.
A measurement at $z\sim0.54$ of
of $\dA/r_d$ using BOSS photometric data was made by \citet{seo12}. 

The first combined 
constraints on  $\dH/r_d$ and $\dA/r_d$ were obtained
using the $z\sim0.3$ SDSS data by  \citet{chuang12} and \citet{xu12}.
Recently,
BOSS has provided precise constraints  on $\dH/r_d$ and $\dA/r_d$
at $z=0.57$  
\citep{anderson13,kazin13}.

At higher redshifts, the BAO feature can be observed using absorption
in the \Lya forest to trace mass, as
suggested by  \citet{macdonald03}, \citet{white03} and \citet{mcdoeisen07}. 
After the observation of the predicted  
large-scale correlations 
in  early BOSS data by \citet{slosar11},
a BAO peak in the \Lya forest correlation function 
was measured by BOSS in the SDSS data release 
DR9 \citep{busca13,slosar13,kirkby13}.
The peak in the quasar-\Lya forest cross-correlation function was
detected in the larger data sets of  DR11 \citep{font13}. 
The DR10 data are now public \citep{dr10ref},
and the DR11 data will be made public simultaneously
with the final SDSS-III data release (DR12) in late 2014.

This paper presents a new measurement of the \Lya forest autocorrelation
function and uses it to study BAO at $z=2.34$.  
It is based on the  methods used by \citet{busca13},
but introduces several improvements in the analysis.
First, and  most important, is a tripling of the number of quasars by using
the DR11 catalog of 158,401 quasars in the redshift range  
2.1~$\leq$~$z_q$~$\leq$~3.5.
Second, to further increase the statistical
power we used a slightly expanded forest range as well as quasars
that have damped \Lya troughs in the forest.
Finally, the \citet{busca13} analysis was based on a decomposition
of the correlation function into monopole  and quadrupole components.
Here, we fit the full correlation $\xi(\rperp,\rpar)$ as a function
of separations perpendicular, $\rperp$, and parallel, $\rpar$, to 
the line of sight.
This more complete treatment is made possible by a more careful
determination of the covariance matrix than was used by \citet{busca13}.

Our analysis uses a fiducial cosmological model
in two places.
First, flux pixel pairs separated in angle and wavelength are assigned
a co-moving separation (in $\hMpc$) using 
the $\dA(z)$ and $\dH(z)$ calculated with
the adopted parameters.
Second, to determine the observed peak position, we compare
our measured correlation function with a correlation function generated
using CAMB \citep{cambref} as described in \citet{kirkby13}.
We adopt the same (flat) \lcdm model
as used in \citet{busca13}, \citet{slosar13}, and \citet{font13};  
with the parameters given in Table~\ref{modeltable}.
The fiducial model has
values of $\dA/r_d$ and $\dH/r_d$ at $z=2.34$ that differ by about 1\%
from the values given by the models favored by CMB data
\citep{planck13,calabrese13} 
given in the second and third columns of 
Table~\ref{modeltable}.

This paper is organized as follows:
Section \ref{samplesec} describes the DR11 data used in this
analysis. Section \ref{mocksec} gives a brief description of the
mock spectra used to test the analysis procedure;
a more detailed description is given in \citet{bautista14}.
Section \ref{xisec} presents our method of estimating the
correlation function $\xi(\rperp,\rpar)$ and its associated
covariance matrix.
In Sect. \ref{fitssec} we fit the data to derive the BAO
peak position parameters, $\dA(z=2.34)/r_d$ and $\dH(z=2.34)/r_d$.
Section \ref{systsec} investigates possible systematic errors
in the measurement.
In Sect. \ref{cosmosec} we compare our measured peak position with that measured by the Quasar-\Lya-forest cross-correlation \citep{font13}
and study \lcdm models that are consistent with these results.
Section \ref{conclusionssec} concludes the paper.

\begin{table}
\begin{center}
\caption{Parameters of the fiducial flat \lcdm cosmological model used
for this analysis, 
the flat \lcdm model derived from Planck and low-$\ell$ WMAP polarization data,
`Planck + WP''  \citep{planck13},
and a flat \lcdm model derived from the 
WMAP, ACT, and SPT data \citep{calabrese13}.
The models are defined by the cold dark matter, baryon, and massive
neutrinos densities, the Hubble constant, and the number of light
neutrino species.
The sound horizon at the drag epoch, $r_d$ is calculated using
CAMB (which can be approximated with
Eq. (55) of \citet{anderson13} to a precision of 0.1\%).
}
\begin{tabular}{l c c c}
  & fiducial & Planck  &WMAP9     \\
  &          &    + WP   & +ACT+SPT \\
\hline \hline
\noalign{\smallskip} 
$\om h^2$  & 0.1323 & 0.14305 & 0.1347 \\
$=\oc h^2$   & 0.1090  & 0.12038 & 0.1122 \\
$\;+\ob h^2$   & 0.0227 & 0.022032 & 0.02252   \\
$\;+\on h^2$    & 0.0006 & 0.0006  & 0 \\
$h $    & 0.7 & 0.6704 & 0.714 \\
$N_\nu$  & 3  & 3 & 3 \\
\hline
\noalign{\smallskip} 
$\om$    &0.27  & 0.3183 & 0.265 \\
$r_d$ (Mpc)   & 149.7 & 147.4 & 149.1  \\
        & (104.80$~h^{-1}$) &  (98.79$~h^{-1}$) & (106.4$~h^{-1}$) \\
$\dA(2.34)/r_d$   & 11.59 & 11.76 & 11.47 \\
$\dH(2.34)/r_d$   & 8.708 & 8.570 & 8.648 \\
\end{tabular}
\end{center}
\label{modeltable}
\end{table}%

\section{BOSS quasar sample and data reduction}
\label{samplesec}

\begin{figure*}[htbp]
\begin{center}
\includegraphics[width=.90\textwidth]{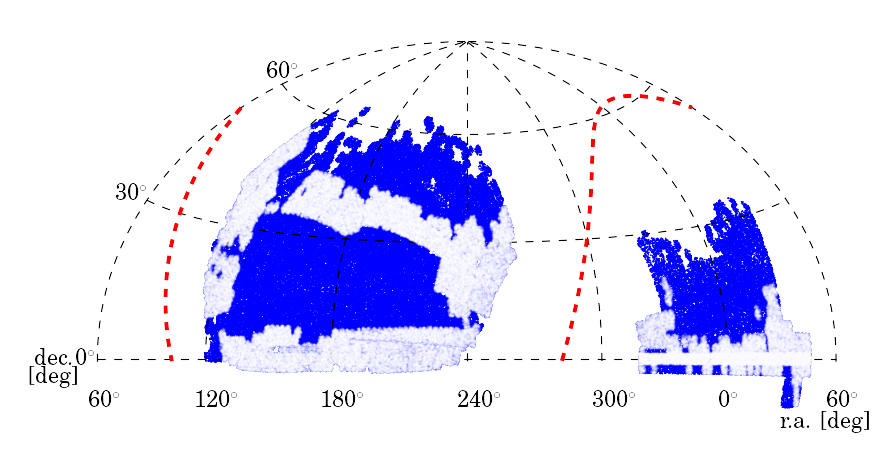}
\caption{Hammer-Aitoff projection of the BOSS DR11 footprint 
(dec. \emph{vs.} r.a.).
The light areas show the DR9 subregion available for the
earlier studies of \citet{busca13} and \citet{slosar13}. The red-dashed
line shows the location of the galactic plane.
}
\label{fig:skysectors}
\end{center}
\end{figure*}

\begin{figure}[htbp]
\begin{center}
\includegraphics[width=.47\textwidth]{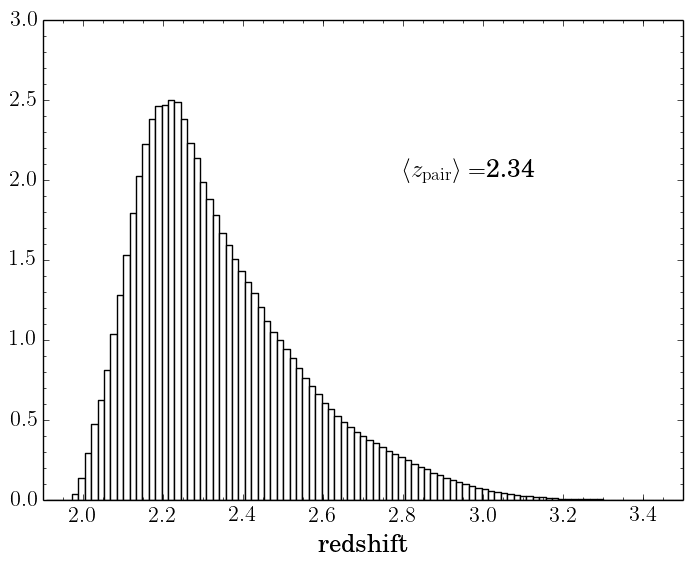}
\includegraphics[width=.47\textwidth]{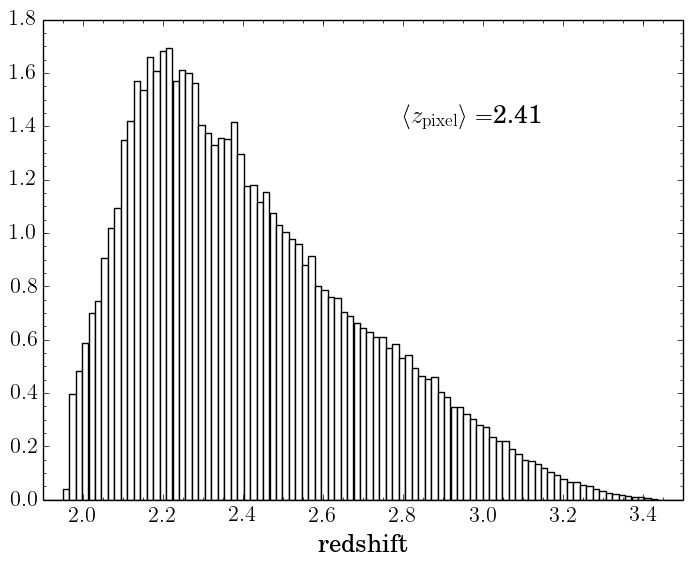}
\caption{Top: redshift distribution of pixel pairs 
contributing to $\xi$ in the region 
$80<r<120~\hMpc$.  Bottom: distribution of all pixel redshifts.}
\label{zdistfig}
\end{center}
\end{figure}

The BOSS project
\citep{bossoverview}
of SDSS-III \citep{eisenstein11}
was designed to obtain the spectra of
over $\sim1.6\times10^6$ luminous galaxies  and  $\sim150,000$ quasars.
The project uses upgraded versions of the
SDSS spectrographs
\citep{bossspectrometer}
mounted on
the Sloan 2.5-meter telescope \citep{gunn06} at Apache Point, New Mexico.

The quasar spectroscopy targets are
selected from photometric data
via a combination of algorithms
\citep{richards09,yeche09,kirkpatrick11,coreref,
palanque10}, as summarized in \citet{ross12}. 
The algorithms use SDSS ugriz fluxes \citep{fuku96,york00}
and, for SDSS Stripe 82, photometric
variability.
Using the techniques of \citet{bovy12},
we also worked with any available
data from non-optical surveys:
the GALEX survey \citep{galexref} in the UV;
the UKIDSS survey \citep{ukidssref} in the NIR,
and the FIRST survey \citep{firstref} in the radio wavelength.

In this paper we use the data from the  DR11 data release
of SDSS-III, whose footprint is  shown in Fig. \ref{fig:skysectors}.
These data cover $8377~{\rm deg^2}$  of the ultimate BOSS
 $10^4~{\rm deg^2}$ footprint.

The data were reduced with the SDSS-III pipeline as described in
\citet{pipeline}.
Typically, four exposures of 15 minutes were co-added in 
pixels of wavelength width 
$\Delta\log_{10}\lambda=10^{-4}$ 
($c\Delta\lambda/\lambda\sim{\rm 69~km\,s^{-1}}$).
The pipeline provides flux-calibrated spectra, object
 classifications (galaxy, quasar, star), and redshift estimates for 
 all targets.

The spectra of all quasar targets were visually inspected
\citep{dr9qso,dr10qso} 
to correct for misidentifications or
inaccurate redshift determinations and to flag broad absorption lines (BALs).
Damped \Lya
troughs were visually flagged, but also identified and characterized
automatically \citep{noterdaeme}. 
The visual inspection of DR11 confirmed
158,401 quasars with 2.1~$\leq$~$z_q$~$\leq$~3.5. 
To simplify the analysis of the \Lya forest, we discarded
quasars with visually identified BALs, leaving 
140,579 quasars. 
A further cut requiring a minimum number of
unmasked forest pixels (50 analysis pixels; see below) 
yielded a sample of 137,562 quasars.

To measure the flux transmission, we used the
rest-frame wavelength interval
\begin{equation}
104.0\,<\lambda_{\rm rf}<120.0~{\rm nm} \; ,
\label{rflambdarange}
\end{equation}
slightly wider than in \cite{busca13}. 
This range is 
bracketed by the \Lyb and \Lya emission lines at 102.5 and 121.6~nm
and was chosen as the maximum range that avoids the large pixel
variances on the slopes of the two lines due to  quasar-to-quasar
diversity of line-emission strength.
The absorber redshift, $z=\lambda/\lambda_{{\rm Ly\alpha}}-1$,
is required to be in the range 
$1.96<z<3.44$. The lower limit
is set by the requirement that the observed wavelength be greater
than 360~nm, 
below which the system throughput is lower than 10\% its peak value. 
The upper limit is produced by the maximum quasar
redshift of 3.5, beyond which the BOSS surface density of quasars
is not high enough to be useful for this study. 
The weighted distribution of redshifts of absorber pairs
near the BAO peak position 
is shown in Fig. \ref{zdistfig} (top panel); 
it has a mean of $\langle z\rangle=2.34$.

Forests with identified DLAs were given a special treatment.
All pixels where the absorption due to the DLA is higher than 20\% were excluded.
Otherwise, the absorption in the wings was corrected  using a 
Voigt profile following
the procedure of \cite{noterdaeme}.  
The metal lines due to absorption at 
the DLA redshift were masked. 
The lines to be masked were identified in a stack of
spectra shifted to the redshift of the detected DLA.
The width of the mask was 0.2~nm or 0.3~nm (depending on the
line strength) or 4.1~nm for \Lyb.
We also masked the $\pm$3~nm region corresponding to \Lya if
the DLA finder erroneously interpreted \Lyb absorption as \Lya
absorption.

We determined the correlation function using analysis pixels 
that are the flux average over three adjacent pipeline pixels.
Throughout the rest of this paper, ``pixel'' refers to analysis
pixels unless otherwise stated.
The width of these pixels is $207~{\rm km~s^{-1}}$, that is, an
observed-wavelength width $\sim0.27\,{\rm nm}$ or $\sim2~\hMpc$.
The total sample of 137,562
 quasars thus provides $\sim 2.4\times10^7$ measurements
of \Lya absorption
over a total volume of $\sim50~h^{-3}{\rm Gpc}^3$.

\begin{figure}[htbp]
\begin{center}
\includegraphics[width=.44\textwidth]{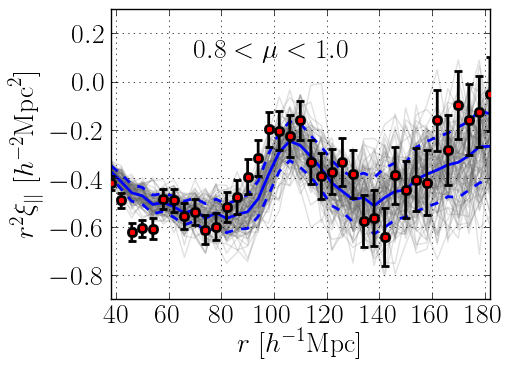}
\includegraphics[width=.44\textwidth]{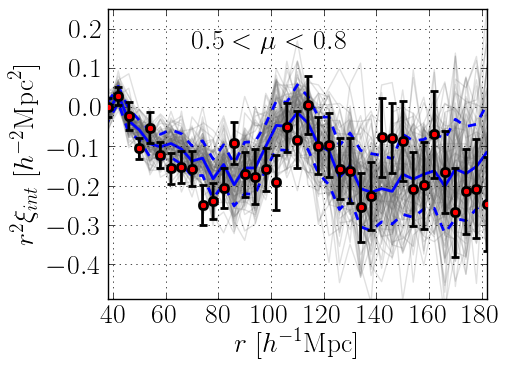}
\includegraphics[width=.44\textwidth]{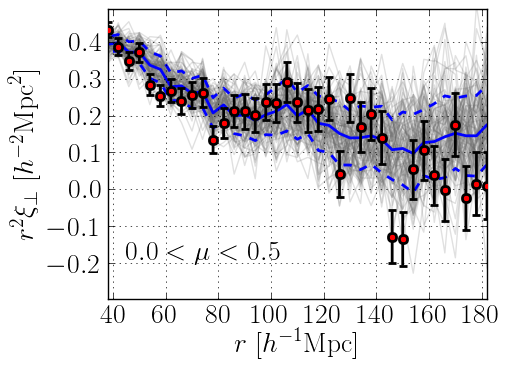}
\caption{Measured 
correlation function averaged over three angular regions:
$\mu>0.8$ (top),  
$0.8>\mu>0.5$ (middle), and 
$0.5>\mu>0.0$ (bottom),
where $\mu$ is the central value of $\rpar/\sqrt{\rpar^2+\rperp^2}$
in each $(\rpar,\rperp)$ bin.
The gray lines show individual sets of mocks, the solid blue line represents
the mean of the 100 mock sets.
The dashed blue lines are the $1\sigma$ variations of the mocks.
The red points show the data.
}
\label{mockdatawedgefig}
\end{center}
\end{figure}

\section{Mock quasar spectra}
\label{mocksec}

In addition to the BOSS spectra, we analyzed 100 sets of mock
spectra.
This exercise was undertaken to search for possible systematic errors in
the recovered BAO peak position and to verify that
uncertainties in the peak position were correctly estimated.
The spectra were generated using the methods of \citet{font12}.
A detailed description of the production and resulting
characteristics of the mock spectra
is given in \citet{bautista14}.

For each set of spectra, the background
quasars were assigned  the angular positions and
redshifts of the DR11 quasars.
The foreground  absorption field in redshift space was generated
according to a cosmology similar to the 
fiducial cosmology of Table \ref{modeltable}.\footnote{
The model has the same $\om h^2$, but $\on h^2=0$.  This change has
a negligible impact on the generated power spectrum and changes
the BAO peak position by only 0.1\%.
}
The unabsorbed spectra (continua)
of the quasars were generated using the principal
component analysis (PCA) eigenspectra of \citet{qsopca},
with amplitudes for each eigenspectrum randomly drawn from Gaussian
distributions with sigma equal to the corresponding eigenvalues as published
in  Table 1 of \citet{suzuki06}.
Finally, the spectra were modified to include the effects of 
the BOSS spectrograph point spread function (PSF), 
readout noise, photon noise, and flux systematic errors.

Our mock production pipeline admits the option of adding DLAs
to the spectra according to the procedure described in \citet{dlasmocks}.
However, since identified DLAs are masked in the analysis of real data, we
did not simulate them into the mocks.  Of course, low column
density and Lyman limit systems are not efficiently identified
and masked in the data,  which means that these systems are present in the data,
but not in the mocks.

Absorption by metals was added to a separate group of ten mocks 
according to the procedure described in \citet{bautista14}.
The quantity of each metal to be added  was determined by 
a modified \Lya stacking procedure 
from \citet{pieri10} and \citet{pieri13}. 
As discussed in Sect. \ref{systsec}, 
the metals have an  effect on the 
recovered correlation function only at small transverse
separations, $\rperp<10~\hMpc$, and have no significant effect 
on the measured position of the BAO peak.

A total of 100 independent metal-free realizations of the BOSS data were 
produced and analyzed with the same procedures as those
for the real data.
Figure \ref{mockdatawedgefig} shows the 
correlation function of the mocks and the data
as measured by the techniques described in the next section.
The mocks reproduce the general features of the observed
correlation function well.
We therefore use them in Sect. \ref{mockanalysissec}
to search for biases in the  analysis procedure that would
influence the position of the BAO peak.

\section{Measurement of the correlation function}
\label{xisec}

\begin{figure}[htbp]
\begin{center}
\includegraphics[width=\columnwidth]{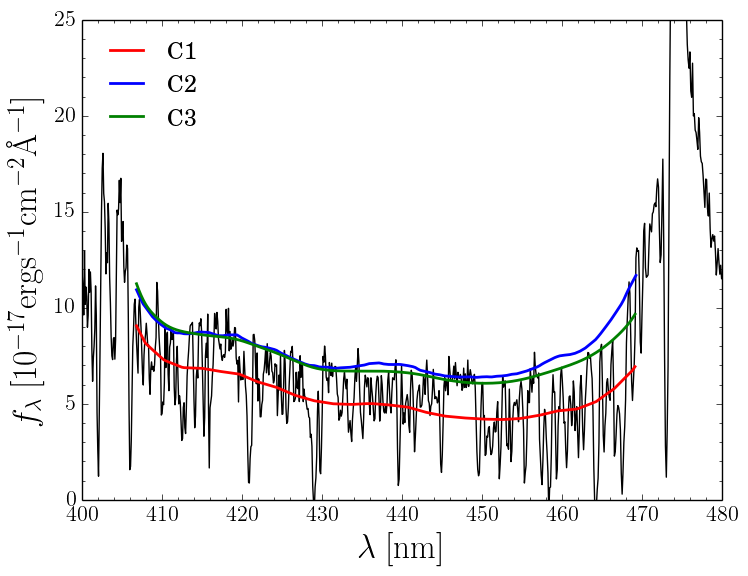}
\caption{
Example of a BOSS quasar spectrum of 
redshift 2.91
The red and blue lines cover the
forest region used here, $104.0<\lambda_{\rm rf}<120.0$.
This region is sandwiched between the quasar's \Lyb and \Lya
emission lines at 
400.9 and 475.4~nm
The blue (green) line is
the C2 (C3) continuum model, $C_q(\lambda)$,
and the red line is the C1 model of the 
product of the continuum and the
mean absorption, $C_q(\lambda)\bar{F}(z)$. (See text.)
}
\label{spectrumfig}
\end{center}
\end{figure}

In this section we describe the measurement of the
correlation function of the transmitted flux fraction:
\begin{equation}
\delta_q(\lambda)={f_q(\lambda)\over C_q(\lambda)\overline{F}(z)}-1 
\; .
\label{delta:def}
\end{equation}
Here, $f_q(\lambda)$ is the observed flux density for quasar $q$
at observed wavelength $\lambda$,
$C_q(\lambda)$ is the unabsorbed flux density 
(the so-called continuum) and
 $\overline{F}(z)$ is the mean transmitted fraction
 at the absorber redshift, $z(\lambda)=\lambda/\lambda_{{\rm Ly\alpha}}-1$. 
Figure \ref{spectrumfig} shows a spectrum with
its $C_q(\lambda)$ (blue line) 
and $C_q\overline{F}$ (red line) estimated by the methods of 
Sect. \ref{continuumsec}.

For the estimator of the  correlation function, we used 
a simple weighted sum of products of the deltas:
 \begin{equation}
 \hat{\xi}_A=\frac{\sum_{ij\in A}w_{ij}\delta_i\delta_j}{\sum_{ij\in A}w_{ij}}
\; ,
\label{eq:thexi}
 \end{equation}
where the $w_{ij}$ are weights (Sect. \ref{weightsec})
and each $i$ or $j$ indexes
a measurement on a quasar $q$ at wavelength $\lambda$.
The sum over $(i,j)$ is understood to run over 
all pairs of pixels 
within 
a bin $A$ in the space of pixel separations, 
$\vec{r}_i-\vec{r}_j$.
The bins $A$ are  defined by a range of width $4~\hMpc$
of the components perpendicular and parallel to the line of sight,
$\rperp$ and $\rpar$.  
We used 50 bins in each component, spanning the range
from $0$ to $200~\hMpc$; the total number of bins used for 
evaluating the correlation function is therefore 2500.
Separations in observational pixel coordinates (ra,dec,$z$) were transformed
into $(\rperp,\rpar)$ in units of $\hMpc$ by using the \lcdm
fiducial cosmology described in Table \ref{modeltable}.

From sum (\ref{eq:thexi}), we excluded
pairs of pixels from the same quasar 
to avoid the correlated errors in $\delta_i$ and $\delta_j$ 
arising from the estimate of $C_q(\lambda)$
for the spectrum of the  quasar.
The weights in Eq. (\ref{eq:thexi}) are set to zero for 
pixels flagged by the pipeline as problematic because of sky emission lines or cosmic rays, for example.
Neither did we use pairs of pixels that had
nearly the same wavelength ($\rpar<4~\hMpc$) and that were
taken on the same focal-plane plate.
The reason for this decision is that these pairs have $\sim20\%$ greater
correlation than expected from our linear
cosmological model fit using data with $\rpar>4~\hMpc$. 
This result is most likely due to spurious correlations introduced by the pipeline, for instance,
from sky subtraction for flux calibration operations.

\subsection{Continuum fits}
\label{continuumsec}

We used three methods to estimate  $C_q\overline{F}$ 
used in Eq. (\ref{delta:def}).
The first two 
assume that 
 $C_q\overline{F}$  is, to first approximation, 
the product of two factors: a scaled  universal
quasar spectrum that is a function of  rest-frame wavelength,
$\lambda_{\rm rf}=\lambda/(1+z_q)$  (for quasar redshift $z_q$), 
and a mean
transmission fraction that slowly varies with absorber redshift.
The universal spectrum is found by
stacking the appropriately normalized spectra of quasars in our
sample,
thus averaging  the fluctuating \Lya absorption.
The continuum
for individual quasars is then
derived from  the universal
spectrum by normalizing it to the quasar's mean forest flux
and then modifying its slope   to account  
for  spectral-index diversity   and/or
photo-spectroscopic miscalibration.

Our simplest continuum estimator, C1, is method 1
of \citet{busca13}. 
It directly estimates the product $C_q\overline{F}$ in Eq.
\ref{delta:def}.
by modeling each spectrum as
\begin{equation}
C_q\overline{F}= a_q \left(
\frac{\lambda}{\langle\lambda\rangle}
\right)^{b_q} \overline{f}(\lambda_{\rm rf},z)
\;,
\label{eq:delta_model_meth1}
\end{equation}
where $a_q$ is a normalization, $b_q$ a deformation parameter, 
$\langle\lambda\rangle$  the mean wavelength 
in the forest
for the quasar $q$, and $\overline{f}(\lambda_\mathrm{rf},z)$ is 
the mean normalized flux  obtained by stacking spectra
in bins of width $\Delta z=0.1$. 

As noted in \citet{busca13},
the mean value of $\delta_q(\lambda)$ (averaged over all measurements
in narrow
bins in $\lambda$) has peaks at the position of
the Balmer series of amplitude $\sim 0.02$.
These artifacts are due to imperfect use of
spectroscopic standards containing those lines.
They are removed on average by subtracting the mean $\delta$
from each measurement:  
$\delta_q(\lambda)\rightarrow\delta_q(\lambda)-\langle\delta(\lambda)\rangle$.

The C1 continuum estimator would be close to optimal if the
distribution of $\delta$ about zero was Gaussian.
Since the true distribution is quite asymmetric, we developed
a slightly more sophisticated continuum estimator, 
method 2 of \citet{busca13},
denoted here as C2.
We adopted this as the standard estimator for this work.
The continuum for each quasar is assumed to be of the form
\begin{equation}
C_q(\lambda)=[a_q + b_q\log(\lambda) ]\overline{C}(\lambda_{\rm rf})
\;,
\end{equation}
where $\overline{C}(\lambda_{\rm rf})$ is the mean continuum  determined
by stacking spectra.
The parameters $a_q$ and $b_q$ 
are fitted to match the quasar's distribution of transmitted flux
to an assumed probability distribution derived from the log-normal model
used to generate mock data.

The C2 continuum is then multiplied by
the 
mean transmitted flux fraction  $\overline{F}(z)$, which we determined by
requiring that the mean of the delta field vanish for all redshifts.
This last step has the effect of removing the average of the Balmer
artifacts.

The third continuum-estimating method, C3,  is a modified version 
of the MF-PCA technique described in \citet{lee12}.
This method has been used to provide continua for the publicly available
DR9 spectra \citep{lee13}.
Unlike the other two methods, it does not assume a universal
spectral form.  
Instead, for each spectrum, 
it fits a variable amplitude PCA template to the part redward of the \Lya
wavelength.
The predicted spectrum in the forest region is then renormalized
so that the mean forest flux matches the mean forest flux at 
the corresponding redshift.

All three methods use data in the forest region to determine the
continuum and therefore necessarily introduce distortions
in the flux transmission field and its correlation
function \citep{slosar11}. 
Fortunately, these distortions are not expected to
shift the BAO peak position, and this expectation is confirmed
in the mock spectra.

\subsection{Weights}
\label{weightsec}

We   chose the weights $w_{ij}$ 
so as to approximately  minimize the relative error on 
$\hat{\xi}_A$ 
estimated with Eq. (\ref{eq:thexi}).
The weights should obviously favor low-noise pixels
and take into account the redshift dependence
of the pixel correlations, 
$\xi_{ij}(z)\propto(1+z_i)^{\gamma/2}(1+z_j)^{\gamma/2}$,
with $\gamma\sim3.8$ \citep{macdonald06}.
Following \citet{busca13}, we used 
\begin{equation}
w_{ij}\propto
{(1+z_i)^{\gamma/2}(1+z_j)^{\gamma/2}\over\xi_{ii}^2\xi_{jj}^2} 
\;,
\label{weights}
\end{equation}
where $\xi_{ii}$ is assumed to have noise and LSS contributions:
\begin{equation}
 \xi_{ii}^2={\sigma^2_{pipeline,i}\over\eta(z_i)}+\sigma^2_{\mathrm{LSS}}(z_i)
\hspace*{5mm}
{\rm and}
\hspace*{5mm}
z_i=\lambda_i/\lambda_{{\rm Ly\alpha}} -1 \;.
\label{twocontris}
\end{equation}  
Here $\sigma_{pipeline,i}^2$
is the pipeline estimate of 
the noise-variance of pixel $i$ multiplied by $(C_i\bar{F_i})^2$,
and $\eta$ is a factor that corrects  for a possibly 
inaccurate estimate of the variance by the pipeline.
The two functions $\eta(z)$ and $\sigma^2_{\mathrm{LSS}}(z)$
are determined by measuring the variance of $\delta_i$ in
bins of $\sigma_{pipeline,i}^2$ 
and redshift.

\subsection{$\xi(\rperp,\rpar)$ and its covariance}
\label{covsec}

\begin{figure}[t]
\begin{center}
\includegraphics[width=.9\columnwidth]{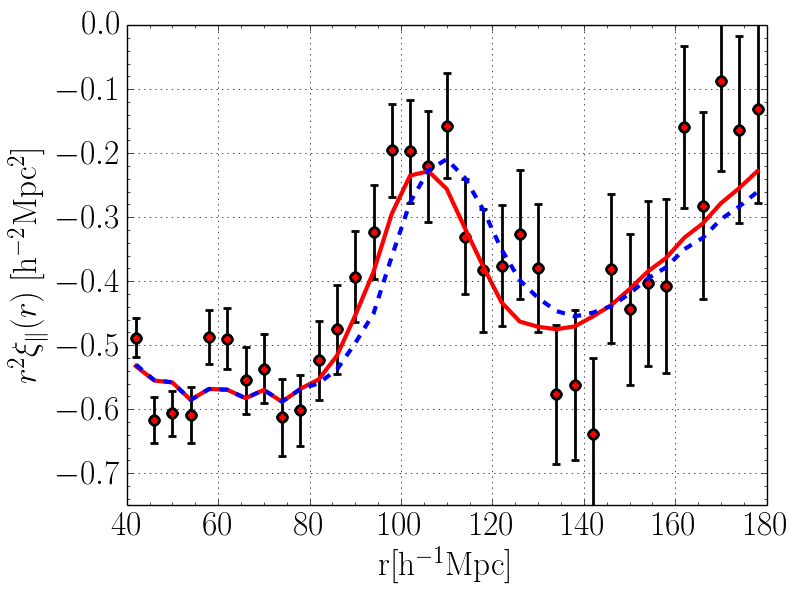}
\includegraphics[width=.9\columnwidth]{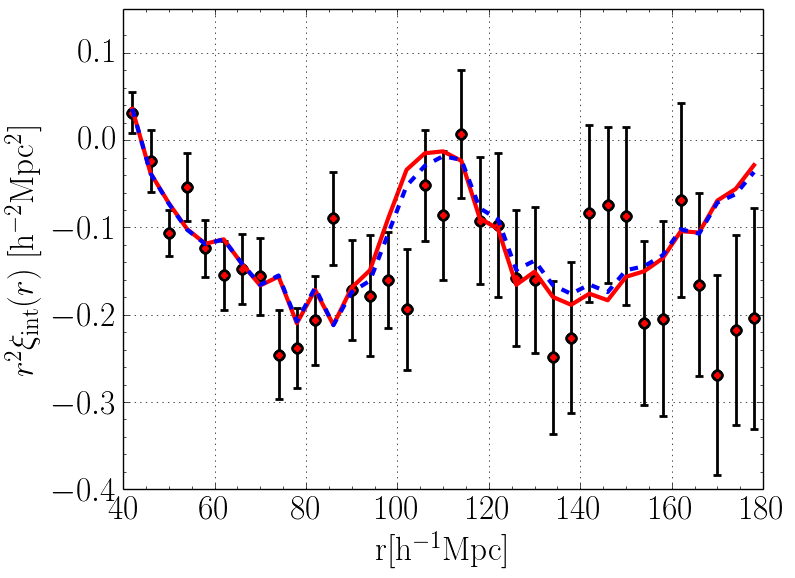}
\includegraphics[width=.9\columnwidth]{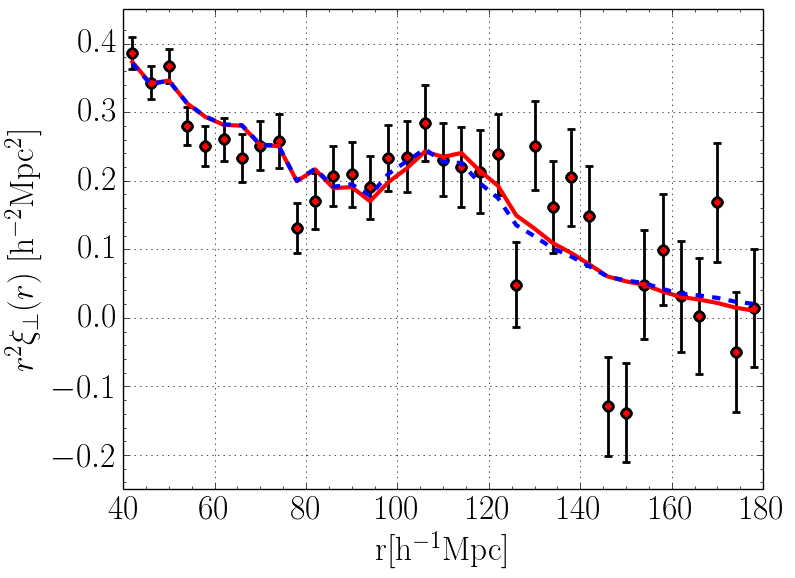}
\caption{
The measured correlation functions (continuum C2) in
three angular regions: 
$\mu>0.8$ (top),  
$0.8>\mu>0.5$ (middle), and 
$0.5>\mu>0.$ (bottom),
where $\mu$ is the central value of $\rpar/\sqrt{\rpar^2+\rperp^2}$
in each $(\rpar,\rperp)$ bin.
The curves show the results of fits as described in Sect. \ref{fitssec}.
The full curve is best fit and the dashed curve is best fit 
when the parameters $\aperp$ and $\apar$ (Eq. \ref{eq:alpha})
are both set to unity.
The irregularities in the fits are due to the use of  
$(\rpar,\rperp)$ bins rather than $(r,\mu)$ bins.
}
\label{wedgesfig}
\end{center}
\end{figure}

The correlation function $\xi(\rperp,\rpar)$ was measured
for the three continuum methods.
Figure \ref{wedgesfig} shows the result using the C2 method,
averaged for three ranges of
$\mu=\rpar/\sqrt{\rpar^2+\rperp^2}$. 
(The analogous plots for C1 and C3 are provided in appendix \ref{wedgeappendix}.)
The superimposed curves present the 
results of fits as described in Sect. \ref{fitssec}. 
The full curve displays the best fit, while the dashed curves 
are the fit when the parameters 
$\aperp$ and $\apar$ (Eq. \ref{eq:alpha})
are set to unity, that is, imposing the BAO peak position of
the fiducial cosmology.

\begin{figure*}[htb] 
\begin{center}
\includegraphics[width=.95\columnwidth]{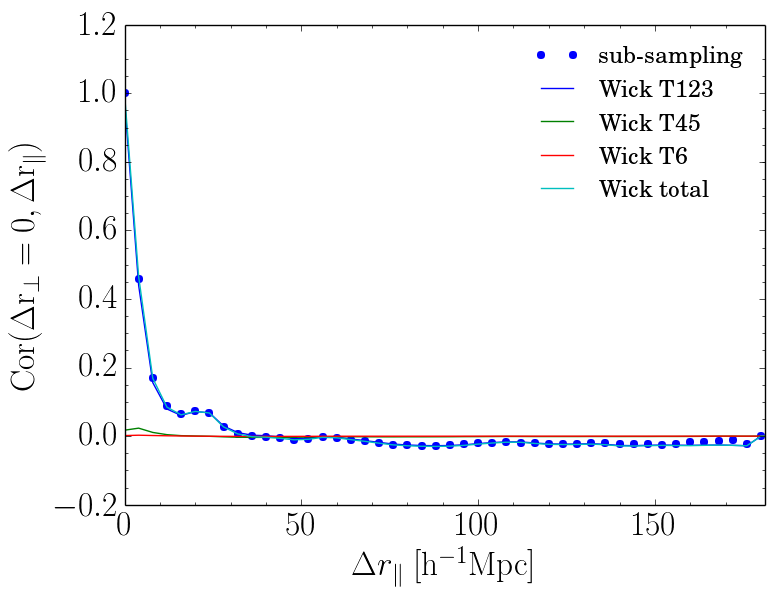}
\includegraphics[width=.95\columnwidth]{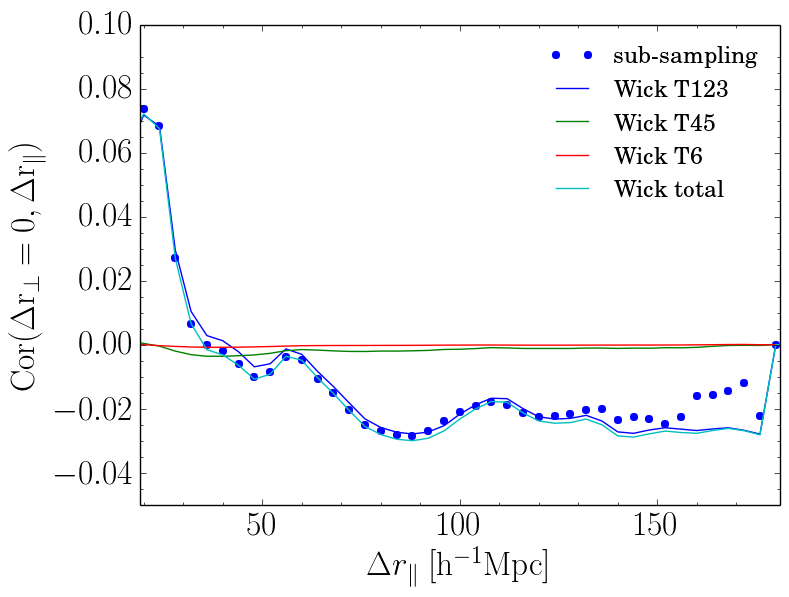}
\includegraphics[width=.95\columnwidth]{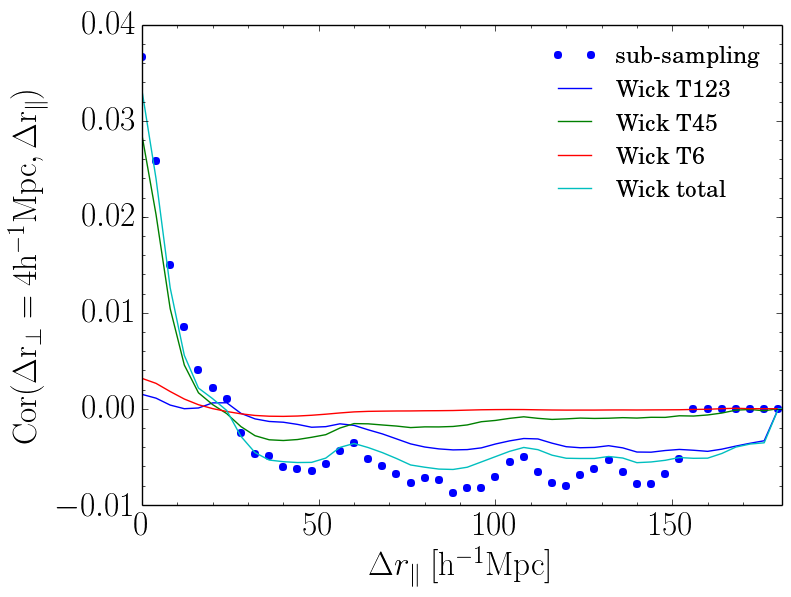}
\includegraphics[width=.95\columnwidth]{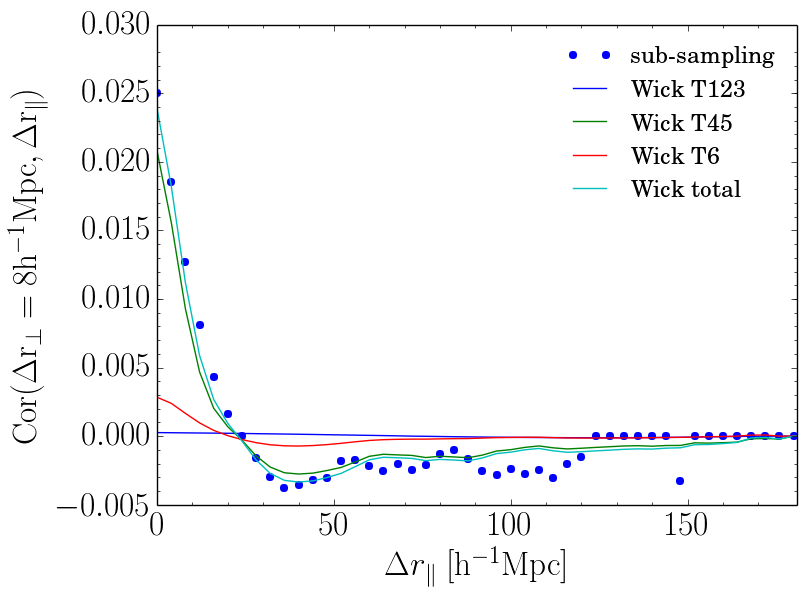}
\end{center}
\caption{
The correlation 
$C(\rperp,\rpar,\rperp^\prime,\rpar^\prime)/
[Var(\rperp,\rpar)Var(\rperp^\prime,\rpar^\prime)]^{1/2}$
as a function of $\rpar-\rpar^\prime$  
(averaged over $(\rperp^\prime,\rpar^\prime)$).
The top figures are
for  $\rperp-\rperp^\prime=0$ over the full range of $\Delta\rpar$ (left)
and for $\Delta\rpar>20\hMpc$ (right).
The bottom two figures are
for  $\rperp-\rperp^\prime=4~\hMpc$ (left) and
for  $\rperp-\rperp^\prime=8~\hMpc$ (right).
Shown 
are the correlations determined by sub-sampling and
by a Wick expansion.  The latter correlations are decomposed into 
the pair-of-pair types, T1-T6, as explained in appendix \ref{covappendix}.
}
\label{drtfig}
\end{figure*}

We evaluated the covariance matrix, 
$C(\rperp,\rpar,\rperp^\prime,\rpar^\prime)$
using two methods described in Appendix \ref{covappendix}.  
The first uses  sub-samples and the second a Wick expansion
of the four-point function of the $\delta$ field.
The two methods give covariances whose differences lead to 
no significant differences in fits for cosmological parameters.
We used the sub-sample covariance matrix in the standard fits.

The $2500\times2500$ element covariance matrix has a relatively 
simple structure.
By far the most important elements are the diagonal elements,
which are, to good approximation, inversely proportional to
the number of pixel pairs used in calculating the correlation
function:
\begin{equation}
C(\rperp,\rpar,\rperp,\rpar) \;\sim\;\frac{0.041}{N_{pairs}}
\; .
\label{varianceeq}
\end{equation}
This is about twice the value that one would calculate assuming that
all pixel pairs used to calculate $\xi(\rperp,\rpar)$ are independent.
This decrease in the effective number of pixels
is due to the correlations between neighboring
pixels on a given quasar:
because of these correlations, a measurement of  $\xi(\rperp,\rpar)$
using a pair of pixels from two quasars
is  not  independent of another measurement of $\xi(\rperp,\rpar)$  
using the same two quasars.

The off-diagonal elements of the covariance matrix also have
a simple structure.
The reasons for this structure are made clear by the Wick expansion
in Appendix \ref{covappendix}, which relates the covariance to 
correlations within pairs-of-pairs of pixels.
The strongest correlations are in pairs-of-pairs where both pairs
involve the \emph{same} two quasars.
To the extent that two neighboring forests are parallel, these
terms contribute only to the covariance matrix elements with
$\rperp=\rperp^\prime$, corresponding to the transverse separation
of the forests.
The elements of the correlation matrix
as a function of $\rpar-\rpar^\prime$ are
illustrated in Fig. \ref{drtfig} (top left);
they closely follow the correlation function $\xi(\Delta\lambda)$
found within individual forests.

The covariance for $\rperp\neq\rperp^\prime$  is due to  
pairs-of-pairs involving three or more quasars and,
for small $|\rperp-\rperp^\prime|$, to the fact that
neighboring forests are not exactly parallel.
As illustrated in Fig. \ref{drtfig}, the covariances are rapidly
decreasing functions of $\rpar-\rpar^\prime$  and
$\rperp-\rperp^\prime$.

The statistical precision of the sub-sampling method is $\sim0.02$ for
individual elements of the correlation matrix.
We adopted this method for the standard analysis because 
it is much faster than the more precise Wick 
method and is therefore better adapted to studies where
the data sample and/or analysis protocol is varied.
Figure \ref{drtfig} shows that only correlations with 
$\Delta\rperp=0,\Delta\rpar<20~\hMpc$ are greater than
the statistical precision and therefore sufficiently large 
for individual matrix elements to be
measured accurately by sub-sampling.
We therefore used the average correlations as a function
of $\rperp-\rperp^\prime$ and $\rpar-\rpar^\prime$, ignoring small
observed variations with $\rperp^\prime$ and $\rpar^\prime$.
The analysis of the mock spectra (Sect. \ref{mockanalysissec})
indicates that this procedure is sufficiently accurate to
produce reasonable $\chi^2$ values 
and that the distribution of estimated BAO peak positions
is similar to that expected from the uncertainties derived from
the $\chi^2$ surfaces.

\section{Fits for the peak position}
\label{fitssec}

To determine the position of the BAO peak in the transverse
and radial directions, we fit the measured $\xi(\rperp,\rpar)$
using the techniques described in \citet{kirkby13}.

\subsection{BAO model}

We fit the measured $\xi(\rpar,\rperp)$ to a form that includes
a cosmological correlation function $\xi_{\rm cosmo}$ and a
broadband function $\xi_{\rm bb}$ 
that takes into account imperfect knowledge of the non-BAO
cosmology and  distortions introduced by the analysis:

\begin{equation}
\xi (\rpar, \rperp,\apar,\aperp)
= \xi_{\rm cosmo} (\rpar,\rperp,\apar,\aperp)
+ \xi_{\rm bb} (\rpar,\rperp) ~.
\label{xitheoryeq}
\end{equation}
The function $\xi_{\rm cosmo}$ is described as a sum of
a non-BAO smooth function and a BAO peak function,
\begin{eqnarray}
\xi_{\rm cosmo}(r_\parallel,r_\perp,\apar,\aperp) = 
\hspace*{40mm}
\cr
\hspace*{10mm}
=\xi_{\rm smooth} (r_\parallel, r_\perp) +
a_{\rm peak} \cdot \xi_{\rm peak} (\apar r_\parallel , \aperp r_\perp ) ~,
\label{xicosmoeq}
\end{eqnarray}
where $a_{\rm peak}$ controls the amplitude of the BAO peak relative
to the smooth contribution.
The radial and transverse dilation factors describing the observed
peak position relative to the fiducial peak position are
\begin{equation}
\apar = \frac { \left[D_H(\bar z)/r_d\right] }{\left[D_H(\bar z)/r_d\right]_{\rm fid}}
\hspace*{3mm}{\rm and}\hspace*{5mm}
\aperp = \frac { \left[D_A(\bar z)/r_d\right] }{\left[D_A(\bar z)/r_d\right]_{\rm fid}} ~,
\label{eq:alpha}
\end{equation}
where $r_d$ is the sound horizon at the drag epoch (defined to sufficient
accuracy for each cosmology by Eq. (55) of \citet{anderson13}).

The function $\xi_{\rm cosmo}$ was calculated from the power spectrum using the following procedure. We modeled the \Lya forest power spectrum including redshift-space distortions and nonlinear effects as
\begin{eqnarray}
P(k,\mu_{k}) = b^{2}(1 + \beta\mu_{k}^{2})^{2}
\hspace*{40mm}
\cr
\hspace*{10mm}
\times\left[ P_{\rm peak}(k)\exp(-k^{2}\Sigma^{2}(\mu_{k})/2) 
+ P_{\rm smooth}(k)\right] ~,
\label{eq:Pk}
\end{eqnarray}
where $\mu_{k} \equiv \hat{z}\cdot\hat{k}$, $b$ is the \Lya forest bias parameter and $\beta$ is the redshift-space distortion parameter. Here, we defined $P_{\rm peak}(k)=P_{\rm lin}(k)-P_{\rm smooth}(k)$, where $P_{\rm lin}$ is the linear-theory matter power spectrum from CAMB \citep{cambref} calculated using the cosmological parameters from the first column of Table \ref{modeltable}, and $P_{\rm smooth}$ is the CAMB power spectrum with the BAO feature erased following the method of \citet{kirkby13}. The exponential function in Eq. (\ref{eq:Pk}) models the anisotropic nonlinear broadening from structure growth \citep{eisenstein07} with $\Sigma^2(\mu_{k})=\mu_{k}^2\Sigma_\parallel^2+(1-\mu_{k}^2)\Sigma_\perp^2$ and is only applied to the BAO feature. The default values we adopted are
$\Sigma_\parallel=6.41~\hMpc$ and
$\Sigma_\perp=3.26~\hMpc$,
which are inferred from the amplitude of the
variation of linear peculiar velocities
along the line of sight that cause a relative displacement
of pixel pairs contributing to the BAO peak form \citep{white14}. 

The power spectrum multipoles are given by
\begin{equation}
P_{\ell}(k) = \frac{2\ell + 1}{2}\, \int_{-1}^{+1} P(k,\mu_k)\, L_{\ell}(\mu_k)\,d\mu_k ~ ,
\end{equation}
where $L_{\ell}$ is the Legendre polynomial. The corresponding correlation function multipoles are then
\begin{equation}
\xi_{\ell,\rm cosmo}(r) = \frac{i^{\ell}}{2\pi^2}\, \int_0^{\infty} k^2 j_{\ell}(k r) \,P_{\ell}(k)\,dk ~ ,
\end{equation}
where $j_{\ell}$ is the spherical Bessel function. Finally, the correlation function is the sum of the multipoles
\begin{equation}
\xi_{\rm cosmo}(r,\mu) = \sum_{\ell=0,2,4}L_{\ell}(\mu)\,\xi_{\ell,\rm cosmo}(r) ~ .
\end{equation}
The nonlinear broadening in principle transfers power to higher even multipoles $\ell=6,8,..$, but the contribution from these higher-order multipoles is negligible.

We wish to ensure the insensitivity of our results to non-BAO cosmology
and to inaccurately modeled astrophysical effects such as UV fluctuations,
nonlinear effects, and DLAs.
We therefore use a broadband function, $\xi_{\rm bb}$ 
to include inaccuracies in the non-BAO
correlation function as well as distortions due, for example, to continuum
fitting.
We used the form
\begin{equation}
\xi_{\rm bb}(r_\parallel,r_\perp) =
\sum_{j=0}^{j_{max}} \sum_{i=0}^{i_{max}} a_{i,l} L_{2j}(\mu)/r^i \;,
\label{xibroadbandeq}
\end{equation}
where the $L_{2j}$ is the Legendre polynomial of order $2j$.
Our standard model uses $(i_{max},j_{max})=(2,2)$.

The standard fits use the fiducial values of
$\Sigma_\perp$ and $\Sigma_\parallel$,
and set $a_{\rm peak}=1$.
They thus have four physical free parameters $(b,\beta,\aperp,\apar)$
and, for the fiducial model, nine broadband distortion parameters.
The standard fit uses the range $40~\hMpc<r<180~\hMpc$, giving
a total of 1515 bins in $(\rpar,\rperp)$ for the correlation
function measurements that are actually used in the fit, and
1502 degrees of freedom.

\subsection{Fits with the mock data sets}
\label{mockanalysissec}

\begin{figure}[htb] 
\includegraphics[width=.95\columnwidth]{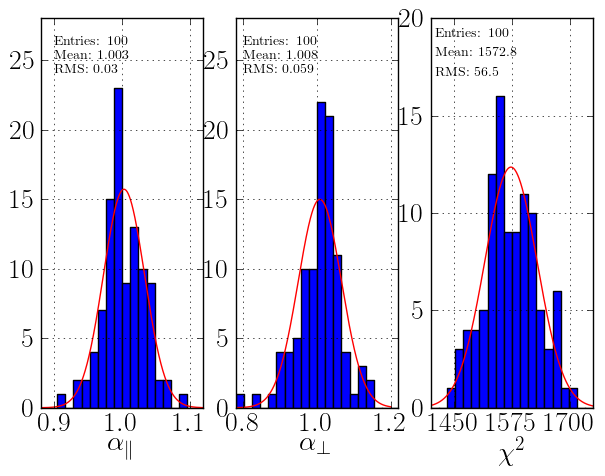}
\includegraphics[width=.95\columnwidth]{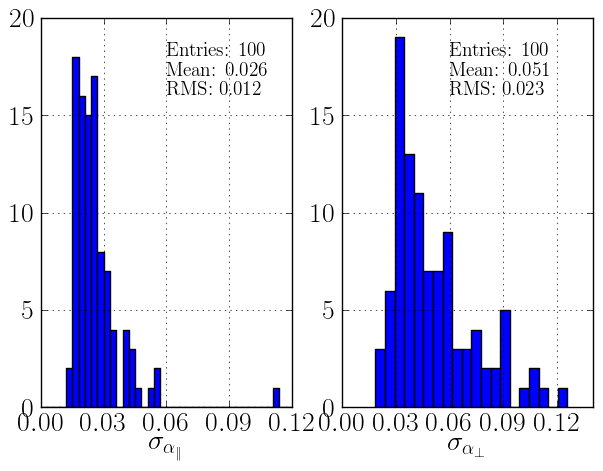}
\caption{
Summary of the results of fits for $(\apar$, $\aperp)$ for
the 100 mock catalogs.
The histograms show
the best-fit values,
the minimum $\chi^2$ values
and the $1\sigma$ uncertainties.
}
\label{aparaperpchi2mocksfig}
\end{figure}

\begin{figure}[htb] 
\includegraphics[width=.95\columnwidth]{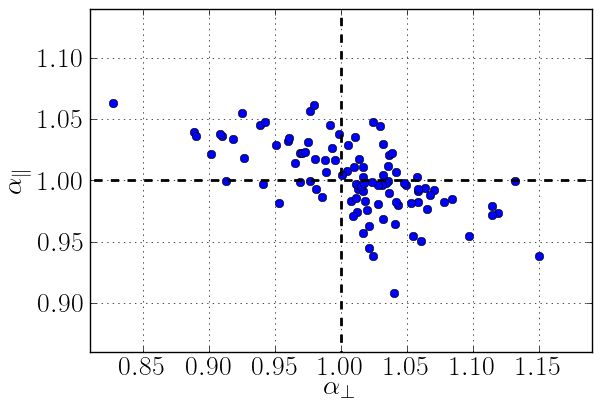}
\caption{Measured $\apar$ and $\aperp$ for the 100 mock catalogs.
}
\label{aparaperpmocksfig}
\end{figure}

\begin{figure}[htbp]
\begin{center}
\includegraphics[width=.95\columnwidth]{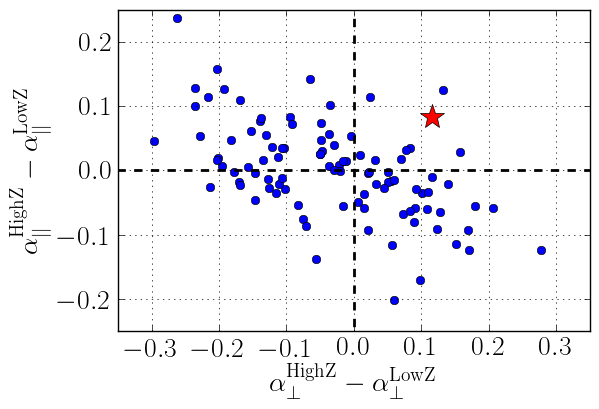}
\caption{
Difference in best-fit $\apar$ and $\aperp$ values between
high redshift ($z>2.295$) and low redshift ($z<2.295$) subsets of 
the 100 mock realizations and the observations (red star).
Compared with Fig. \ref{aparaperpmocksfig}, the plot shows
the degraded precision resulting from dividing the data into two
redshift bins.
}
\label{hilozcompfig}
\end{center}
\end{figure}

Figure \ref{aparaperpchi2mocksfig}
summarizes the results of the cosmological fits on the 100 sets of mock spectra.
The mean recovered $\apar$ and $\aperp$ 
are consistent with unity,
indicating no bias in the measurement of the BAO peak
position.\footnote{
After we analysed the mocks, we realized that they had
been analyzed with a model (the fiducial model of Table \ref{modeltable})
that was slightly different from the model used to produce them, which
had the same $\om h^2$, but $\on h^2=0$ instead of $\on h^2=0.0006$.
Its value of  $r_d$ is 0.15~Mpc lower than the fiducial $r_d$,
so the expected mean values of $\aperp$ and $\apar$ are 0.999, which
is sufficiently close to unity for the precision of this study.
}
The numbers of $\apar$ measurements within $1\sigma$ and $2\sigma$ of unity
are 61 and 93, consistent with the expected numbers, 68 and 95.5.
For $\aperp$ the numbers are 68 and 95.
For the combined $(\apar,\aperp)$ measurements,
70 are within the $1\sigma$  and  93 within the $2\sigma$ contours.

The r.m.s. deviations  of $\apar$ and $\aperp$ 
are $0.029$ and $0.057$.
The mean $\chi^2$ is similar to the number of degrees of freedom, 
indicating that the model represents the mock observations sufficiently
well and that the covariance matrix is well estimated.

Figure \ref{aparaperpmocksfig} shows an anticorrelation  between 
the recovered
$\apar$ and $\aperp$, with a correlation coefficient of $-0.6$.
The quantity of the form $\apar^w\aperp^{1-w}$ with the
smallest mock-to-mock variance 
has $w\sim0.7$, with an r.m.s. deviation of 0.017.
This result is to be compared with the optimal quantity for galaxy
surveys, $\sim\apar^{1/3}\aperp^{2/3}$.  The difference arises because redshift distortions are stronger
for the \Lya forest,
a consequence of the low bias factor that
enables more precise measurements in the $\rpar$ direction even though
there are  two dimensions for $\rperp$ and only one for $\rpar$.
This effect is evident in Fig. \ref{wedgesfig} where the
BAO peak is most easily seen for $\mu>0.8$.

Figure \ref{hilozcompfig} presents the results of fits 
separating the mock data into two redshift
bins, $z<2.295$ and $z>2.295$.
The differences between 
the measured $\apar$ and $\aperp$ for the two bins
are typically of about 10\%.

\subsection{Fits with the observations}

\begin{table*}
\begin{center}
\caption{
Results for the standard fit and modified fits.
The standard fit uses the C2 continuum, 
a broadband defined by $(i_{\rm max},j_{\rm max})=(2,2)$,
a forest defined by $104.0<\lambda_{\rm rf}<120.0~{\rm nm}$,
and $a_{\rm peak}=1$.
}
\begin{tabular}{l c c c c c}
analysis  
    & $\apar$ 
    & $\aperp$ 
    & $\beta$ 
    & $b(1+\beta)$
    & $\chi^2_{\rm min}/DOF$  \\
\noalign{\smallskip} 
\hline
\noalign{\smallskip} 
standard(C2)
    & $1.054^{+0.032}_{-0.031}(1 \sigma)^{+0.069}_{-0.063}(2 \sigma) $ 
    & $0.973^{+0.056}_{-0.051}(1 \sigma)^{+0.199}_{-0.103}(2 \sigma) $ 
& $1.50\pm 0.47$
& $ -0.402 \pm 0.024$
&   1499.1/(1515-13) \\
\noalign{\smallskip} 
\hline
\noalign{\smallskip} 
C1 
    & $1.038^{+0.035}_{-0.037}(1 \sigma)^{+0.073}_{-0.074}(2 \sigma) $
    & $1.054^{+0.132}_{-0.093}(1 \sigma)^{+0.246}_{-0.176}(2 \sigma) $
& $3.47 \pm 2.78 $
& $-0.43 \pm 0.06 $
&  1571.5/(1515-13) \\
\noalign{\smallskip} 
C3
    & $1.038^{+0.026}_{-0.039}(1 \sigma)^{+0.054}_{-0.071}(2 \sigma) $
    & $1.041^{+0.259}_{-0.063}(1 \sigma)^{+0.259}_{-0.126}(2 \sigma) $ 
& $2.28 \pm 1.25 $
& $-0.48 \pm 0.05 $
&  1603.8/(1515-13) \\
\noalign{\smallskip} 
$\beta$-prior $(1.4\pm0.4)$ 
    & $1.055^{+0.032}_{-0.031}(1 \sigma)^{+0.068}_{-0.063}(2 \sigma) $
    & $0.972^{+0.053}_{-0.051}(1 \sigma)^{+0.117}_{-0.102}(2 \sigma) $ 
& $1.41 \pm 0.34 $
& $-0.40 \pm 0.04 $
& 1499.1/(1515-13)\\
\noalign{\smallskip} 

$a_{\rm peak}$ free
    & $1.054^{+0.035}_{-0.031}(1 \sigma)^{+0.078}_{-0.063}(2 \sigma) $ 
    & $0.973^{+0.057}_{-0.052}(1 \sigma)^{+0.232}_{-0.104}(2 \sigma) $ 
& $1.50 \pm 1.10 $
& $-0.39 \pm 0.25 $
&   1499.0/(1515-14)\\
\noalign{\smallskip} 

$\Sigma_\perp=\Sigma_\parallel=0$
    & $ 1.053^{+0.029}_{-0.028}(1 \sigma)^{+0.062}_{-0.059}(2 \sigma)$
    & $ 0.961^{+0.055}_{-0.052}(1 \sigma)^{+0.254}_{-0.103}(2 \sigma)$
& $1.30 \pm 0.80 $
& $-0.35 \pm 0.05 $
&  1501.2/(1515-13)\\
\noalign{\smallskip} 

$\Sigma_\perp,\Sigma_\parallel$ free
    & $1.063^{+0.041}_{-0.036}(1 \sigma) ^{+0.101}_{-0.073}(2 \sigma)$ 	
    & $0.976^{+0.053}_{-0.05}(1 \sigma)^{+0.124}_{-0.102}(2 \sigma)$	
& $ 0.50 \pm 0.40$   
& $ -0.42 \pm 0.06$
&  1495.7/(1515-15) \\
\noalign{\smallskip}

no special DLA treatment
    & $1.049^{+0.038}_{-0.034}(1 \sigma)^{+0.089}_{-0.068}(2 \sigma)$ 
    & $0.954^{+0.053}_{-0.049}(1 \sigma)^{+0.132}_{-0.096}(2 \sigma)$	
& $0.36 \pm 0.46 $  
& $ -0.34 \pm 0.06 $
&  1489.7/(1515-13) \\ 
\noalign{\smallskip} 

\hline
\noalign{\smallskip} 

104.5$<\lambda_{\rm rf}<$118.0~nm
    & $1.052^{+0.041}_{-0.041}(1 \sigma)^{+0.145}_{-0.094}(2 \sigma) $
    & unconstrained 
& $2.37 \pm 2.81 $  
& $ -0.35 \pm 0.08 $
&  1448.2/(1515-13)\\

\noalign{\smallskip} 

No spectra with DLAs 
    & $1.031^{+0.035}_{-0.035}(1 \sigma)^{+0.074}_{-0.074}(2 \sigma) $
    & $1.073^{+0.117}_{-0.082}(1 \sigma)^{+0.228}_{-0.171}(2 \sigma) $
& $2.38 \pm 1.93$
& $ -0.44 \pm 0.06 $
&  1506.5/(1515-13)\\
\noalign{\smallskip}

\hline
\noalign{\smallskip} 
$z<2.295$
    & $0.996^{+0.052}_{-0.054}(1 \sigma)^{+0.113}_{-0.134}(2 \sigma) $ 
    & $0.89^{+0.064}_{-0.053}(1 \sigma)^{+0.148}_{-0.108}(2 \sigma) $ 
& $ 1.10 \pm 0.94 $
& $ -0.32 \pm 0.07 $
&  1523.3/(1515-13)\\
\noalign{\smallskip} 

$z>2.295$
    & $1.096^{+0.037}_{-0.036}(1 \sigma)^{+0.079}_{-0.073}(2 \sigma) $ 
    & $0.994^{+0.057}_{-0.049}(1 \sigma)^{+0.155}_{-0.1}(2 \sigma) $ 
& $1.61 \pm 1.05 $ 
& $ -0.50 \pm 0.06$
&  1479.1/(1515-13)\\

\noalign{\smallskip} 

\hline
\noalign{\smallskip} 
$a_{\rm peak}=0$ 
    & - 
    & -
    & -
    & - 
    & 1526.2/(1515-11)  
\label{dlytable}
\end{tabular}
\end{center}
\end{table*}%

\begin{figure}[tb] 
\includegraphics[width=.95\columnwidth]{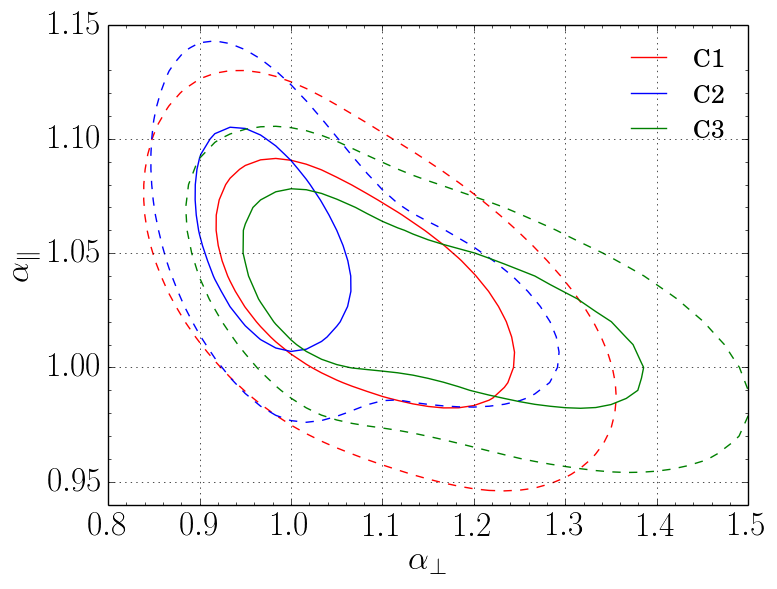}
\caption{
Constraints on $(\apar,\aperp)$ using the three
continuum estimators, C1 (red), C2 (blue), and C3(green).
The solid and dashed contours correspond to
$1\sigma$ and $2\sigma$   ($\Delta\chi^2=2.3,6.2$).
}
\label{aparaperpdatafig}
\end{figure}

Table \ref{dlytable} gives the results of fits of the data
for a variety of data sets and analysis assumptions.
The first line lists our standard analysis using the C2 continua:
\begin{equation}
\apar=1.054^{+0.032}_{-0.031}
\hspace*{5mm}{\rm and}\hspace*{5mm}
\aperp=0.973^{+0.056}_{-0.051}
\; .
\end{equation}
The precisions on $\apar$ and $\aperp$ 
inferred from our $\chi^2$ fitting procedure 
are typical of those found using the 100 mock catalogs
(Fig. \ref{aparaperpchi2mocksfig}).
The full contours presented in Fig. \ref{aparaperpdatafig}
show that these errors are somewhat non-Gaussian, with an
anti-correlation between $\apar$ and $\aperp$.
In particular, the $2\sigma$ contour extends asymmetrically to 
large $\aperp$, consistent with the visual impression from
Fig. \ref{wedgesfig}.
The most precisely determined combination is
\begin{equation}
\apar^{0.7}\aperp^{0.3} = 1.025 \pm 0.021 \;.
\end{equation}

The next seven lines of Table \ref{dlytable} 
present the results of analyses using the
standard data set, but with modified  assumptions:
using the non-standard continua C1 and C3;
adding a Gaussian prior to the redshift distortion parameter 
around its nominal value $\beta=1.4$ of width $0.4$;
freeing the peak amplitude $a_{\rm peak}$; 
fitting the nonlinearity parameters, $\Sigma_\parallel$ and $\Sigma_\perp$,
or 
setting them to zero (and thus not correcting for nonlinearities);
using spectra with one or more DLA, but without a
special treatment (fit with Voigt profile).
Because these seven fits all use the same data set, any variation of the
results at the $1\sigma$ level might indicate a systematic
effect.  In fact, all configurations produce results that are
consistent at the sub-$\sigma$ level.
We note, however, that 
the $1\sigma$ precision on $\aperp$ is degraded through
use of C1 and C3, although C1 does almost as well as C2 at the $2\sigma$ level. 
The higher sensitivity of the $\aperp$ uncertainty to the
continuum method compared
with that for the $\apar$ uncertainty
might be expected from the low statistical
significance of the BAO peak in transverse directions.
The central value and error of $\apar$ are robust
to these differences.

The next two lines in 
table \ref{dlytable} are the results for C2 
with reduced data sets:  
using a short forest (104.5$<\lambda_{\rm rf}<$118.0~nm)
farther away from the \Lya and \Lyb peaks, or
removing spectra with one or more DLAs.
Both results are consistent at $1\sigma$ with those 
obtained with the more aggressive
standard data set 
but, as expected, with larger statistical errors.

The next two lines present the results obtained by dividing
the pixel-pair sample into two redshift bins.
The two results now correspond to fairly independent samples
and agree at the 2$\sigma$ level.
The differences between measurements
of $(\apar,\aperp)$ for the two redshift bins in the mock spectra
are displayed in Fig. \ref{hilozcompfig},  suggesting
that the observed difference is similar to that observed with the mock spectra.

The last line of Table \ref{dlytable} is the $\chi^2$ for a fit
without a BAO peak.  Comparison with the first line reveals $\Delta\chi^2=27.2$
for two additional degrees of freedom, corresponding to a 5$\sigma$ detection.

\section{Systematic errors}
\label{systsec}

\begin{figure}[tb] 
\includegraphics[width=.9\columnwidth]{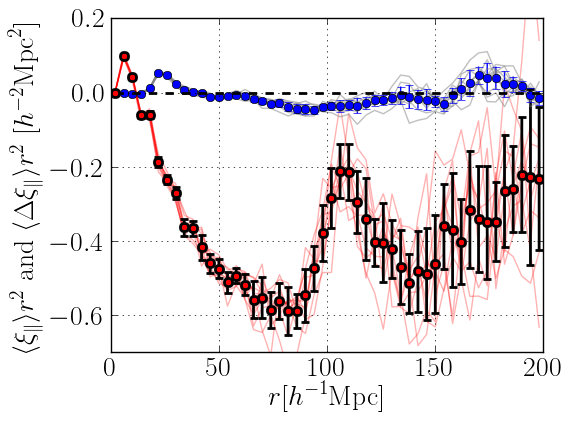}
\caption{
Effect of metals on the measured correlation function for
10 mock sets.
The the red circles show $r^2\xi(r)$ for $\mu>0.8$ 
averaged over the 10 mock sets
The blue circles show the difference between $r^2\xi(r)$ and $r^2\xi(r)$
in the same mock realization, but without metals.
The light red and blue lines show the results for individual mock
sets, the error bars give the standard deviation of the 10
realizations.
}
\label{metalwedgefig}
\end{figure}

\begin{figure}[htb] 
\includegraphics[width=.9\columnwidth]{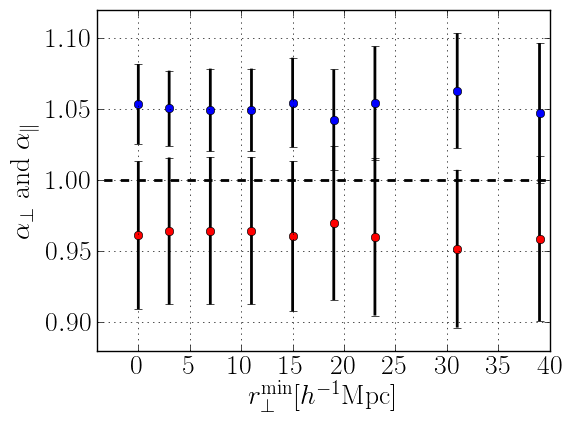}
\caption{
Values $\apar$ (blue dots) and $\aperp$ (red dots)
recovered from the DR11 data  
for different choices of the minimum
transverse separation, $r_\perp^{\rm min}$, used in the fit.
The measured values do not change significantly when eliminating
the small $\rperp$ bins that may be contaminated by correlations
due to absorption by metals.
}
\label{rtcutfig}
\end{figure}

\begin{table}
\begin{center}
\caption{
Fit results with the C2 continuum with a modified fitting range
[standard: $40<r<180~\hMpc$] and number of terms in the
broadband (Eq. \ref{xibroadbandeq})  
[standard:$(i_{\rm max},j_{\rm max})=(2,2)$].
}
\begin{tabular}{l c c c}
  & $\apar$ & $\aperp$ &
$\chi^2_{\rm min}/DOF$  \\
\noalign{\smallskip} 
\hline 
\noalign{\smallskip} 
standard (C2) &$1.054\pm0.032$  & $0.973\pm0.055$  &1499.0/(1515-13)   \\
\noalign{\smallskip} 
\hline
range  &  &  &  \\
$\;\;\;(\hMpc)$  &  &  &  \\
$60<r< 180$ & $1.045\pm0.032$ & $0.986\pm0.063$ & 1391.8/(1415-13) \\
$40<r< 160$ & $1.052\pm0.033$  & $0.974\pm0.053$ & 1139.2/(1177-13) \\
\hline
\noalign{\smallskip} 
$(i_{\rm max},j_{\rm max})$ & & & \\
(2,3) & $1.057\pm0.032$ & $0.970\pm0.050$ & 1484.2/(1515-16) \\
(3,2) & $1.050\pm0.033$ & $0.987\pm0.067$ & 1497.8/(1515-16) \\
(3,3) & $1.051\pm0.034$ & $0.986\pm0.068$ & 1479.2/(1515-20) \\
\label{systable}
\end{tabular}
\end{center}
\end{table}

The uncertainties reported in Table \ref{dlytable} are 
statistical and are derived from the $\chi^2$ surface.
In this section we discuss possible systematic errors.
We find no evidence for effects that add uncertainties similar to the statistical errors.

We derived cosmological information 
by comparing  the measured flux correlation
function with a model, defined
by Eq. (\ref{xitheoryeq}), (\ref{xicosmoeq}) and (\ref{xibroadbandeq}),
that depends on cosmological 
parameters (Eq. \ref{xicosmoeq}).
Systematic errors in the derived parameters
can result if either the assumed model
or the measured correlation function differ systematically
from the true flux-correlation function.

In fitting the data, we added a general broadband
form $\xi_{\rm bb}$ (Eq. \ref{xibroadbandeq}) to the 
assumed cosmological correlation function
$\xi_{\rm cosmo}$.
The role of $\xi_{\rm bb}$ is to  make the fit sensitive only
to the position of the BAO peak and not to the more uncertain
smooth component of $\xi_{\rm cosmo}$.  
To verify that the broadband does indeed remove any sensitivity
to smooth components of the correlation function, 
we varied the form of the assumed broadband and the range
over which it was fit.  
The results, listed in Table \ref{systable}, show
no significant variation of the derived $(\apar,\aperp)$, 
indicating that the broadband performed as required.
Of particular significance, adding greater freedom to
$\xi_{\rm bb}$ only has an impact of about $10\%$ impact on the 
size of the $\apar$ error, although it has a stronger
impact $(20-30\%)$ on the $\aperp$ error.

Because the use of $\xi_{\rm bb}$ makes our results insensitive
to smooth features in $\xi$, 
we are primarily concerned with rough effects either due to
observing or analysis artifacts or  to
physical effects that invalidate the assumed 
theoretical form (Eq. \ref{xitheoryeq}).

We first considered errors in 
the theoretical form of the correlation function. 
The function $\xi_{\rm cosmo}$ in (\ref{xitheoryeq})
is subject to uncertainties arising from nonlinear
effects and, more importantly, in the astrophysical processes that 
determine the flux transmission correlations from matter correlations.
The resulting  uncertainties in the dominant
\Lya absorption would be expected to generate only errors
that vary slowly with $\vec{r}$ and are therefore absorbed
into $\xi_{\rm bb}$.
On the other hand, 
absorption by metals, not included in (\ref{xitheoryeq}), generates an 
excess correlation in individual forests in the form of narrow peaks centered
on the wavelength separations between
$\lambda_{\rm Ly\alpha}$ and metal lines.
For example, the SiII(1260.42) absorption correlated with \Lya(1215.67)
absorption 
gives rise
to a narrow peak of excess correlation at $r=110~\hMpc$ on
the line of sight, at $z=2.34$. This narrow peak is smeared because of
the range of observed redshifts, over a width $\Delta r \sim \pm 5~\hMpc$.
This correlation in the absorption in individual quasar spectra induces
a correlation in the spectra of neighboring quasars
(small $\rperp$), because they probe
correlated structures of \Lya and SiII absorption in the intergalactic medium.

As described in \citet{bautista14}, we added metals
to the mock spectra to estimate their importance.
As expected,
the mocks
indicate that the metal-induced correlation rapidly decreases
with transverse separation, 
dropping by a factor 
of five between the first and third $\rperp$ bin. 
Figure \ref{metalwedgefig} 
shows the effect on ten sets of mocks in the important region $\mu>0.8$.
The modifications do show structure that might not be modeled
by our broadband term.  
Fortunately, the effect 
has no significant impact on the measured position of
the  BAO peak, with $(\apar,\aperp)$ 
for  metaled and metal-free mocks having a
mean difference  and mock-to-mock dispersion of 
$\Delta\apar=0.002\pm0.003$ and
$\Delta\aperp=0.003\pm0.009$.

Because the amplitude of the metal lines is somewhat uncertain,
we empirically verified 
that they are unimportant by re-performing the fit of the DR11 data
after excising the correlation function bins with $\rperp<r_\perp^{\rm min}$.
Most of the metal correlation occurs within $\rperp<10~\hMpc$, 
so any unexpected effect from metal lines
would have to be made apparent by a dependence of the BAO results
on $r_\perp^{\rm  min}$.  
In fact, the results
are remarkably stable, as shown in Fig. \ref{rtcutfig}.
We thus conclude that absorption by metals is unlikely to significantly
affect the measured position of the BAO peak.

We now examine artifacts introduced by the analysis.
The measured correlation function will be different
from the true flux-transmission correlation function
because of systematic errors in the flux-transmission field, 
$\delta_q(\lambda)$,
defined by Eq. (\ref{delta:def}).
Such errors can be
introduced through an inaccurate flux-calibration 
or an inaccurate estimate 
of the function  $C_q(\lambda)\bar{F}(z)$.
These errors in $\delta_q(\lambda)$ will generate systematic errors
in the correlation function if the neighboring quasars have
\emph{correlated} systematic errors.

The most obvious error in the $\delta_q(\lambda)$ arises
from the necessity of using a spectral template to estimate
the continuum.
The C1 and C2 methods use
a unique template that is multiplied by a linear function
to fit the observed spectrum.  
This approach results in two systematic errors on the correlation function.
First, as previously noted, fitting a linear function for the 
continuum effectively
removes broadband power in individual spectra.
This error will be absorbed into $\xi_{\rm bb}$ and not
generate biases in the BAO peak position.
Second, $\delta_q(\lambda)$ along individual lines
of sight will be incorrect because non-smooth quasar spectral diversity
is not taken into account by the universal template.
However,
because the peculiarities of individual quasar spectra are
determined by local effects, they would not be expected to be
correlated between neighboring quasars. 
The tests with the mock catalogs that include uncorrelated spectral diversity
confirm that the imprecisions of the continuum estimates 
do not introduce biases into the estimates of the BAO peak positions.

Errors introduced by
the flux calibration are potentially more dangerous.
The BOSS spectrograph \citep{bossspectrometer} is calibrated by observing
stars whose spectral shape is known.
Most of these objects  are F-stars whose spectra contain
the Balmer series of hydrogen lines.
The present BOSS pipeline procedure for calibration
imperfectly treats the standard spectra in the neighborhood
of the Balmer lines, resulting in calibration vectors, $C(\lambda)$, that
show peaks at the Balmer lines of amplitude 
$\langle \Delta C/C\rangle\sim0.02\pm0.004$, where the $\pm0.004$
refers to our estimated quasar-to-quasars r.m.s. 
variation of the Balmer artifacts \citep{busca13}.
If uncorrected, these calibration errors would lead to
$\delta\sim 0.02$ at absorber redshifts corresponding
to the Balmer lines.
The
subtraction of the mean $\delta$-field, $\langle\delta(\lambda)\rangle$
in our analysis procedure removes this effect on average,
but does not correct calibration vectors individually.
Because of the relative uniformity of the Balmer feature in 
the calibration vectors, this mean correction is 
expected to be sufficient.
In particular, we have verified that no significant changes
in the  correlation
function appear when it is calculated taking into account 
the observed correlations in  the Balmer artifacts, $\Delta C/C$.

To verify this conclusion, we searched for Balmer artifacts in 
the measured $\xi(\rpar,\rperp,\langle\lambda\rangle)$
where $\langle\lambda\rangle$ is the mean wavelength of the
pixel pair.
If our mean correction is insufficient, there would be
excess correlations at $\rpar=0$ and $\langle\lambda\rangle$
equal to a Balmer wavelength.
Artifacts  would also appear at $\rpar$ and $\langle\lambda\rangle$
corresponding to pairs of Balmer lines.
For example, the pair 
[
H$\delta$ (410~nm),
H$\epsilon$ (397~nm)]
would produce excess correlation at the corresponding
radial separation $98~\hMpc$ and $\langle\lambda\rangle=403$~nm.
A search has yielded no significant correlation excesses.
Additionally, removing from the analysis pixel pairs near (397,410)nm,
dangerously near the BAO peak,
does not generate any measurable change in the BAO
peak position.

\section{Cosmological implications}
\label{cosmosec}

\begin{figure}[tb] 
\includegraphics[width=.9\columnwidth]{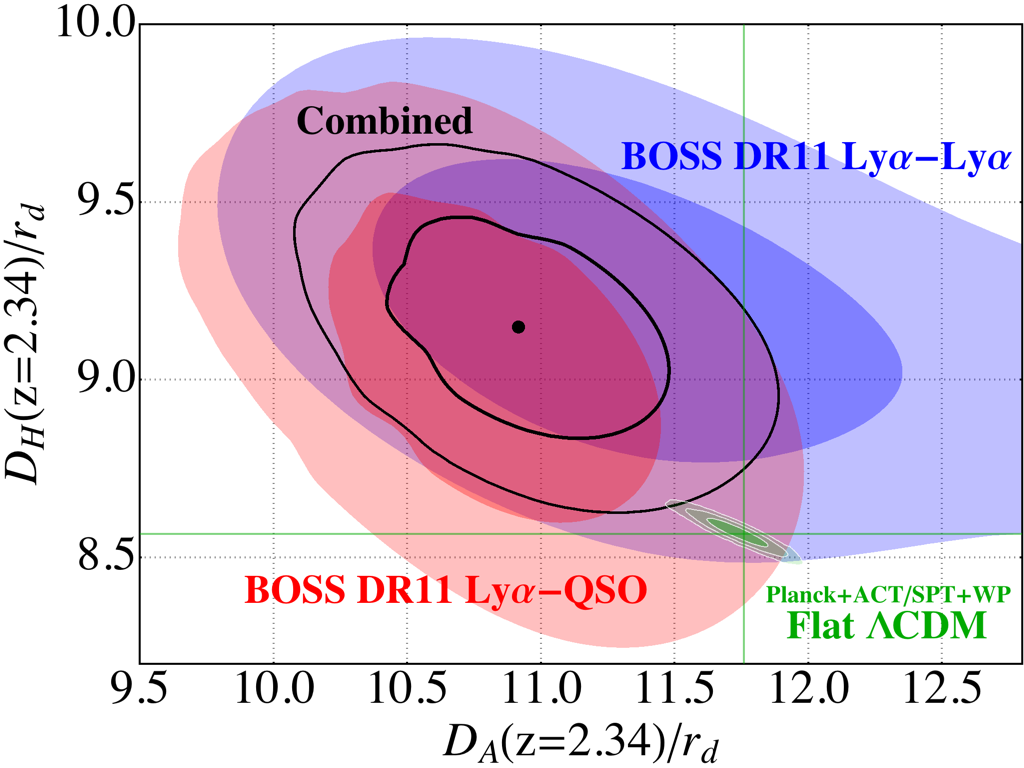}
\caption{
Constraints on $(\dA/r_d,\dH/r_d)$. Contours show 68.3\% ($\Delta\chi^2=2.3$)
and 95.5\% ($\Delta\chi^2=6.2$)
contours from the \Lya forest autocorrelation (this work, blue), 
the quasar \Lya forest cross-correlation
\citep{font13} (red), and the combined constraints (black).
The green contours are CMB constraints 
calculated using the Planck+WP+SPT+ACT chains
\citep{planck13} assuming a flat \lcdm cosmology.
}
\label{crossautofig}
\end{figure}

\begin{figure}[htb] 
\includegraphics[width=.9\columnwidth]{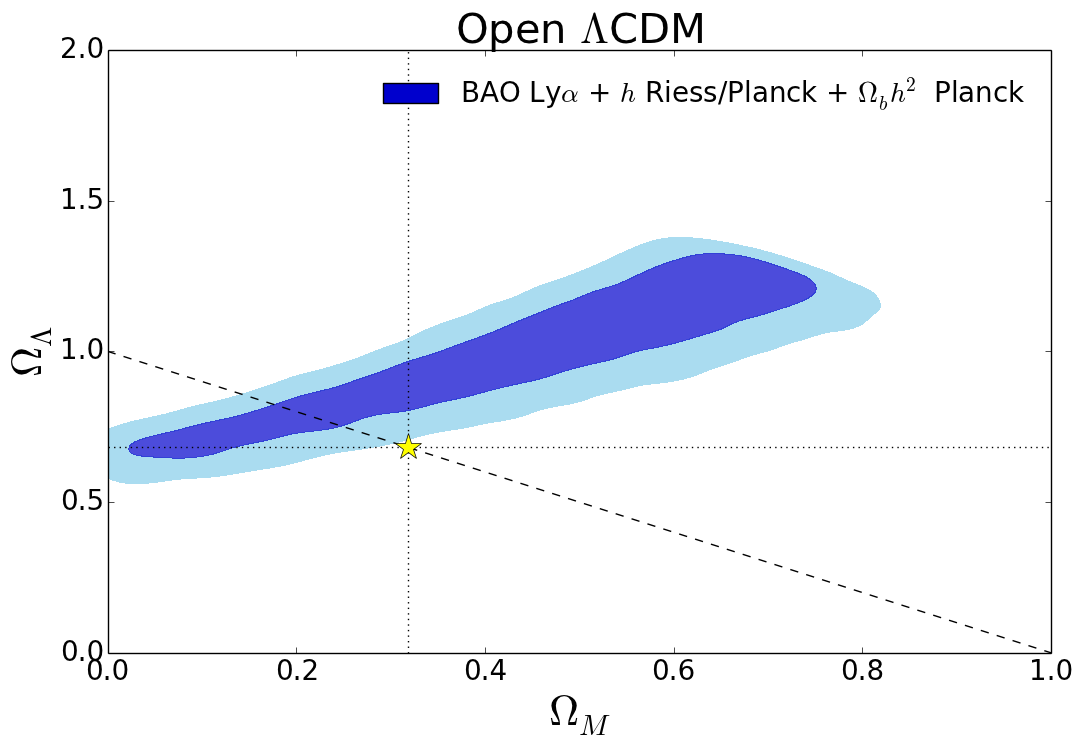}
\caption{Constraints
on the o\lcdm parameters  $(\ol,\om)$
based on the autocorrelation 
contours of Fig. \ref{crossautofig}.
The contours show 68.3\% and 95.5\% confidence levels. 
The Planck value of $\ob h^2$ is assumed together with a Gaussian prior
for $H_0=70.6\pm3.2~{\rm km~s^{-1}Mpc^{-1}}$.
The yellow star is the Planck \lcdm measurement, the dashed line
corresponds to a flat Universe.
}
\label{olcdmfig}
\end{figure}

The standard fit values for $(\apar,\aperp)$ 
from Table~\ref{dlytable} combined with the fiducial
values from Table \ref{modeltable} yield the following results:
\begin{equation}
\frac{\dH(2.34)}{r_d}
                     = 9.18 \pm 0.28(1\sigma) \;\pm0.6(2\sigma)
\label{dhconstraint}
\end{equation}
and
\begin{equation}
\frac{\dA(2.34)}{r_d}
                    = 11.28 \pm0.65(1\sigma) \; ^{+2.8}_{-1.2}(2\sigma)
\; .
\label{daconstraint}
\end{equation}
The blue shading in Fig.~\ref{crossautofig} shows the 68.3\% and 95.5\% 
likelihood contours for these parameters, which are mildly anticorrelated.  
These constraints can be expressed equivalently as
\begin{eqnarray}
H(z=2.34) &=& (222 \pm 7 \hubunits)\times \frac{147.4\,{\rm Mpc}}{r_d} \cr
\dA(z=2.34) &=& (1662 \pm  96\,{\rm Mpc}) \times  \frac{r_d}{147.4\,{\rm Mpc}},
\label{HandDresultseq}
\end{eqnarray}
where we have scaled by the value $r_d = 147.4$ Mpc from the Planck+WP model
in Table~\ref{modeltable}.

Our measured values of $\dH/r_d$ and $\dA/r_d$  (Eq. \ref{dhconstraint}
and \ref{daconstraint})  
can be compared with those predicted by 
the two  CMB inspired flat \lcdm models from 
Table~\ref{modeltable}: $(8.570,11.76)$ for Planck+WP and 
$(8.648,11.47)$ for WMAP9+ACT+SPT.
Figure~\ref{crossautofig} demonstrates that our values differ by 
$1.8\sigma$ from those of  the Planck+WP model.  They differ from
the  WMAP9+ACT+SPT model by  $1.6\sigma$.
We emphasize that, in contrast to the values of $\alpha_\parallel$
and $\alpha_\perp$, the constraints quoted in 
Eq.~(\ref{dhconstraint})-(\ref{HandDresultseq}) are independent
of the fiducial model adopted in the analysis, at least over a
substantial parameter range.  We have confirmed this expectation
by repeating some of our analyses using the Planck+WP parameters
of Table~\ref{modeltable} in place of our standard fiducial model,
finding negligible change in the inferred values of $\dH/r_d$
and $\dA/r_d$.

To illustrate this tension, we show in Fig. \ref{olcdmfig}
values of $\om$ and $\ol$ 
that are consistent with our measurements.
We consider models of \lcdm with curvature, having four free
parameters: the cosmological constant, the matter and the baryon
densities, and the Hubble constant
$(\ol,\om,\ob h^2,h)$.
For clarity, 
we  placed priors on two of them, adopting the
Planck value of $\ob h^2$ \citep{planck13} and adopting
a wide prior on
$h=0.706\pm0.032$ 
meant to cover the value measured with the local distance ladder 
\citep{riess2011} and that measured with CMB anisotropies
assuming
a \lcdm cosmology \citep{planck13}.

With these priors,
Fig. \ref{olcdmfig}  shows 
that the flat \lcdm~model preferred by CMB data \citep{planck13}
lies near the 95.5\% confidence level of our measurement.
We note that values of $(\om,\ol)$ far from the the line $\om+\ol=1$
disagree with the combination of CMB data and BAO at $z<1$,
which require $1-\om-\ol=-0.0005\pm0.0066$ \citep{planck13}. 


The tension with CMB data is also seen in the
BAO measurement using the  quasar-\lya\ forest
cross-correlation \citep{font13}.

The measurement of this function follows a procedure that is similar
to that used here, except that the estimator (\ref{eq:thexi})
is replaced with $\sum w_i\delta_i/\sum w_i$ where the sum is now
over forest pixels $i$ separated from any quasar within a 
range of $(\rperp,\rpar)$.

Red contours in Fig.~\ref{crossautofig} show the 68.3\% and 95.5\%
likelihood contours derived from the cross-correlation.
The implied values of $D_A$ and $D_H$ are consistent between
the autocorrelation and cross-correlation measurements, but
the statistical errors are interestingly complementary.
The autocorrelation constrains $D_H$ more tightly than $D_A$ 
because redshift-space distortions are so strong in \lya\ forest
clustering, a consequence of the low bias factor of the forest.
While there are far fewer quasar-forest pairs than forest-forest
pairs, the cross-correlation still yields a useful BAO signal because the
quasars themselves are highly biased, which boosts the clustering
amplitude.  However, redshift-space distortions are weaker in the
cross-correlation for the same reason, and the cross-correlation
analysis yields comparable statistical errors in the transverse
and line-of-sight BAO.  The cross-correlation constraint therefore
suppresses the elongated tails of the autocorrelation likelihood
contours seen in Fig.~\ref{aparaperpdatafig} toward high
$D_A$ and correspondingly low $D_H$.

The statistical errors in these BAO measurements are dominated
by combinations of limited sampling of the volume probed,
by instrumental noise in the
\lya\ forest spectra, and (for the cross-correlation measurement)
by shot noise of the quasar density field.
For this reason, the statistical errors in the two BAO 
measurements are almost completely uncorrelated, as discussed
 in detail in Appendix~\ref{crosscovariance}.
We therefore combined the two likelihood surfaces as
if they were independent to produce the joint likelihood contours shown by
the solid lines in Fig.~\ref{crossautofig}.
Marginalized 1-D constraints from the combined likelihood are
\begin{equation}
\frac{\dH(2.34)}{r_d} = 
9.15_{- 0.21}^{ + 0.20}\, (1\sigma)\,
_{- 0.42}^{ + 0.40}\,(2\sigma)
\label{dhcombined}
\end{equation}
and
\begin{equation}
\frac{\dA(2.34)}{r_d} =
10.93_{ - 0.34}^{ + 0.35}\, (1\sigma)\,
_{- 0.65}^{ + 0.75}\,(2\sigma)
\; .
\label{dacombined}
\end{equation}

These numbers can be compared with the 
green contours in Fig.~\ref{crossautofig}, 
which show the 68.3\%, 95.5\%, and 99.7\%
confidence contours on $D_A/r_d$ and $D_H/r_d$ derived from
CMB data (specifically, using the Planck + WMAP polarization + SPT + ACT
chains available from the Planck Collaboration; 
this data set
gives results very similar to the Planck+WP model of Table \ref{modeltable}),
assuming a flat \lcdm cosmological model.
These predictions lie outside the 95.5\% likelihood interval for
the combined cross- and autocorrelation BAO measurements, 
an $\approx 2.5\sigma$ tension with the data.
The tension with the WMAP9+ACT+SPT model 
of Table \ref{modeltable} is slightly smaller, $\approx 2.2\sigma$.
In more detail, the Planck \lcdm prediction 
is approximately $2\sigma$
below the value of $D_H$ inferred from the autocorrelation and
approximately $2\sigma$ above the value of $D_A$ inferred from
the cross-correlation,
deviations of $\approx 7\%$ in each case.
The tension between the CMB-constrained flat \lcdm model and the
autocorrelation measurement of $D_H$ is evident in the top
panel of Fig.~\ref{wedgesfig}, 
where the peak in the data is visually to the left of 
the peak in the fiducial model (and would be even more to the left
of the Planck+WP model).

How seriously should one take this tension?
For the autocorrelation function,
the success of our method in reproducing the correct parameters
when averaged over our 100 mock catalogs, 
and the insensitivity of our derived $\alpha_\parallel$ and $\alpha_\perp$
to many variations of our analysis procedure as discussed in
\S\ref{systsec}, both suggest that systematic biases in 
our measurements should be 
smaller than our quoted errors.
The agreement between the directly estimated statistical errors
and the scatter in best-fit $\alpha$ values for our mock catalogs
indicate that our error estimates themselves are accurate,
although with 100 mock catalogs we cannot test this accuracy
stringently.  The most significant impact seen with varying
the analysis procedure in \S\ref{systsec} is the larger statistical
errors on $\alpha_\parallel$ and $\alpha_\perp$ for the continuum
subtraction method C1.  Detailed examination of the likelihood
contours in Fig.~\ref{aparaperpdatafig} shows that the
larger $\alpha_\parallel$ errors (and lower central value)
for the C1 method are 
a consequence of its weaker constraint on $\alpha_\perp$,
which allows contours to stretch into the region of 
low $\alpha_\parallel$ and high $\alpha_\perp$.
These regions are inconsistent with the stronger $\alpha_\perp$
constraints of the cross-correlation measurement, so in a joint
likelihood they would be eliminated in any case.
Furthermore, our standard C2 subtraction method is clearly
more realistic than the C1 method because it is based on a
more realistic flux PDF rather than a Gaussian approximation to it.
Nonetheless, the variation seen in Table~\ref{dlytable}
suggests some degree of caution about the precision of
our statistical errors, even though we have no clear
evidence that they are underestimated, particularly because
of the relatively weak detection of the transverse BAO signal.

We note that while we used mock spectra to verify the 
statistical errors of the autocorrelation function, this
was not possible for the cross-correlation function
because we do not have mock spectra with \lya~absorption
correlated with quasar positions.
We are in the process of producing such mock spectra, and
they will be used in future publications.

While it is premature to conclude that a major modification
of \lcdm is needed, it is nevertheless interesting
to note what sorts of changes are indicated by the data.
The most widely discussed extensions to flat \lcdm, allowing nonzero
space curvature or a dark energy equation-of-state with $w \neq -1$,
do not readily resolve the difference seen in Fig.~\ref{crossautofig}
without running afoul of other constraints.
This is because of the necessity of 
decreasing $\dA(2.34)$ while
increasing $\dH(2.34)$,
which is difficult because the former
is related to the integral of the latter.

Requirements for more general forms of dark energy can be found by
considering our measurement of $H(z)$, which, combined with the Friedman
equation, determines the density of dark energy $\rhode(z)$.
Assuming space to be flat and matter to be conserved, and neglecting the
radiation density, 
we have
\begin{equation}
  {8\pi G \over 3} \rhode(z) = H^2(z) - H_0^2 \om (1+z)^3 ~.
\label{rhodeofzeq}
\end{equation}
Dividing by $\rhode(z=0)$ gives
\begin{equation}
\frac{\rhode(z)}{\rhode(z=0)}
=\frac{H(z)^2-\om H_0^2(1+z)^3}{(1-\om) H_0^2} ~.
\end{equation}
The uncertainty on $\rhode(z=2.34)/\rhode(0)$
is dominated by the difference between
two nearly equal numbers in the numerator. If we use
the precise values of $r_d$ and $\om h^2=0.143\pm0.003$ from 
the Planck+WP CMB power spectrum measurement, 
the uncertainty is dominated by that of
our value of $r_dH(z=2.34)$ (eqn. \ref{dhcombined}).
We find
\begin{equation}
\frac{\rhode(z=2.34)}{\rhode(z=0)}\,=\,
-1.2 \pm 0.8 ~.
\end{equation}
The difference of $\sim 2.5\sigma$ from the expected value of unity
for the \lcdm model is the same as the difference discussed above (although
the quoted error above implies a $2.8\sigma$ deviation, this is
reduced slightly by the non-Gaussianity of the likelihood distribution
of the measured $D_H$). If
a negative value of $\rhode$ were to persist as measurement errors
on $H(z)$ from BAO and $\om h^2$ from the CMB
are improved, this would imply 
that the dark energy
density at z=2.4 is lower than that of z=0, perhaps even with the
opposite sign.
This conclusion could be avoided if
matter were not conserved from the
epoch of recombination,  
(invalidating the use of the Planck value of $\om h^2$ in Eq.
\ref{rhodeofzeq}),
or that the Universe is closed 
(adding a positive term to the r.h.s.
of Eq. \ref{rhodeofzeq}).
If, in addition to explaining the low value
of $H(z=2.34)$, one wishes to  reproduce the low observed
value of $\dA(z=2.34)$, there are further constraints on the model. 
For example, a flat dark-energy model  that \emph{lowers}
the value of $\rhode(z=2.34)$ to
reproduce the observed $H(z=2.34)$ would need to \emph{increase}
$\rhode$ for $0.7<z<2.0$ so as to decrease the value of $\dA(z=2.34)$
while maintaining $\dA(0.57)$.
A compensating \emph{decrease} of $\rhode(z>2.34)$
would then be needed to maintain the observed value of $\dA$ at the
last-scattering surface.
Detailed discussions of such models will be presented in
a forthcoming publication (The BOSS collaboration, in preparation).

\section{Conclusions}
\label{conclusionssec}

The \Lya correlation data presented in this study
constrain $\dH/r_d$ and $\dA/r_d$
at $z\sim2.34$.\footnote{
The baofit software used in this paper is publicly available at 
http://github.com/deepzot/baofit/. 
The measured cross-correlation function and its covariance matrix, 
and the instructions to reproduce the main BAO results presented 
in this paper, can be downloaded from
 http://darkmatter.ps.uci.edu/baofit/, together with the 
likelihood surfaces used to generate the contours in Fig. \ref{crossautofig}.
}
The 3.0\% precision on $\dH/r_d$ and 5.8\% precision on $\dA/r_d$ obtained
here improve on the precision
of previous measurements: 8\%  on  $\dH/r_d$ \citep{busca13},
and 3.4\% on $\dH/r_d$ and 7.2\% on $\dA/r_d$ \citep{slosar13}. 
The increasing precision of the three studies is primarily
due to their increasing statistical power, rather than to 
methodological improvements.
The 2\% precision on the optimal combination $\dH^{0.7}\dA^{0.3}/r_d$
can be compared with the 1\% precision for $D_V(z=0.57)/r_d$
obtained by \citet{anderson13}.

The derived values of $\dH/r_d$ and $\dA/r_d$
obtained here with the
\Lya\ autocorrelation are similar to those inferred
from the Quasar-Ly$\alpha$-forest cross-correlation \citep{font13},
as shown in Fig. \ref{crossautofig}.
At the two-standard-deviation level,
the two techniques are separately compatible with the Planck+WP
and fiducial models of Table~\ref{modeltable}.
However, the combined constraints are inconsistent with the 
Planck+WP \lcdm model at $\approx 2.5\sigma$ significance,
given our estimated statistical uncertainties.
The tests presented in earlier sections suggest that our
statistical error estimates are accurate and that systematic
uncertainties associated with our modeling and analysis procedures
are smaller than these statistical errors.  

We are in the process of addressing what we consider to be
the main weaknesses of our analysis.
The  
artifacts in the spectrophotometric calibration due, for example, to
Balmer lines, will be eliminated.
More sophisticated continuum modeling making use of spectral features
will allow us to verify that unsuspected correlated continua in neighboring
quasars are not introducing artifacts in the autocorrelation function.
Finally, we are producing realistic mock catalogs with quasar
positions correlated with \lya~absorption features in the
corresponding forests.
Such mocks would allow us to verify the statistical errors
for the cross-correlation measurement and to search for
unsuspected correlations between the cross- and autocorrelation
function measurements.
All of these improvements in the analysis procedure
will be used for publications using the higher statistical
power of the upcoming DR12 data release.


The cosmological implications of our results will be investigated in
much greater depth in a forthcoming paper
(The BOSS collaboration, in preparation), where we combine the Ly$\alpha$-forest 
BAO with the BOSS galaxy BAO results at lower redshift and with
CMB and supernova data, which enables interesting constraints on a
variety of theoretical models.

\begin{acknowledgements}

Funding for SDSS-III has been provided by the Alfred~P. Sloan Foundation, the Participating Institutions, the National Science Foundation, and the U.S. Department of Energy Office of Science. The SDSS-III web site is http://www.sdss3.org/.

SDSS-III is managed by the Astrophysical Research Consortium for the Participating Institutions of the SDSS-III Collaboration including the University of Arizona, the Brazilian Participation Group, Brookhaven National Laboratory, Carnegie Mellon University, University of Florida, the French Participation Group, the German Participation Group, Harvard University, the Instituto de Astrofisica de Canarias, the Michigan State/Notre Dame/JINA Participation Group, Johns Hopkins University, Lawrence Berkeley National Laboratory, Max Planck Institute for Astrophysics, Max Planck Institute for Extraterrestrial Physics, New Mexico State University, New York University, Ohio State University, Pennsylvania State University, University of Portsmouth, Princeton University, the Spanish Participation Group, University of Tokyo, University of Utah, Vanderbilt University, University of Virginia, University of Washington, and Yale University.

The French Participation Group of SDSS-III was supported by the
Agence Nationale de la
Recherche under contracts ANR-08-BLAN-0222 and ANR-12-BS05-0015-01.

Timoth\'ee~Delubac and Jean-Paul~Kneib acknowledge support from the ERC advanced grant LIDA.

Matthew~Pieri has
received funding 
from the European Union Seventh Framework Programme (FP7/2007-2013) 
under grant agreement n° [PIIF-GA-2011-301665].

The authors acknowledge the support of France Grilles 
for providing computing resources on the French National 
Grid Infrastructure.

\end{acknowledgements}

\appendix
\section{Covariance matrix}
\label{covappendix}

\subsection{Estimation of the covariance via a Wick expansion.}

The Wick  expansion for the covariance between $\xi$ for
two bins $A$ and $B$ is the  
sum over pairs of pairs:
\begin{equation}
C_{AB}\; = 
S_{AB}^{-1}
\sum_{i,j\in A} \sum_{k,l\in B}
w_i w_j w_k w_l
[\xi_{ik}\xi_{jl} + \xi_{il}\xi_{jk}]
\end{equation}
with
\begin{equation}
\xi_{ij} \;=\; \langle \delta_i \delta_j \rangle
\;.
\end{equation}
The pairs $(i,j)$ and $(k,l)$ refer to the ends of
the vectors $r_A\in A$ and $r_B\in B$.
The normalization factor is
\begin{equation}
S_{AB} \;=\; \sum_{i,j\in A} (w_i w_j) \sum_{k,l\in B} (w_k w_l)
\; .
\end{equation}

\begin{figure}[htb] 
\includegraphics[width=.47\textwidth]{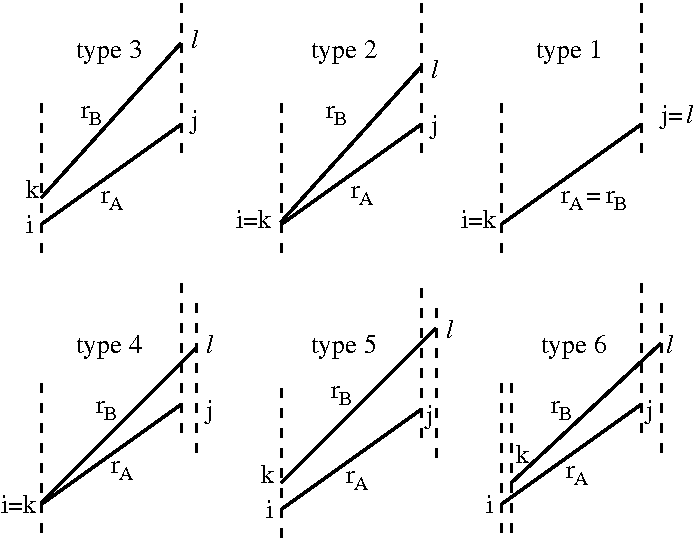}
\caption{Six types of pairs of pairs.  The dashed lines refer
to the quasar lines-of-sight.
The variances are dominated by types 1,2, and 3.
The ($\rperp-\rperp^\prime=0$) covariances are dominated
by types 2 and 3.}
\label{typesfig}
\end{figure}

As illustrated in Fig. \ref{typesfig}, 
there are six types of  pairs-of-pairs,
$(ijkl)$,
characterized by the number of distinct 
points (2,3,4) and numbers of quasars (2,3,4).

The complete sum of pairs-of-pairs would require a prohibitively
long computer time.  We therefore evaluated the
sum by using only a  random sample of pairs-of-pairs
and by 
replacing products of distinct pixels with the previously 
evaluated correlation function, 
either 1D for pairs involving only one quasar, or 
3D for pairs involving two quasars

The variances, Eq. \ref{varianceeq}, 
are dominated ($\sim97$\%) by the two-quasar diagrams in 
Fig. \ref{typesfig}. About 60\% 
of the variance is produced by  the diagonal
diagram $(i=k,\; j=l)$.  
The nondiagonal terms, 
$(i=k,\; j\ne l)$ and $(i\ne k,\; j\ne l)$ account for 25\% and 15\% of the variance.
The dominant covariances, that is, those with $\rperp=\rperp^\prime$
and $0<\rpar-\rpar^\prime<20~\hMpc$, are dominated by
the nondiagonal two-quasar diagrams.

The Wick results for the important covariance matrix elements
are summarized in Fig. \ref{drtfig}.

\subsection{Estimation of the covariance via sub-sampling.} 

We used a sub-sample method to estimate the covariance matrix. 
The method consists of 
organizing the space of pairs of quasars into \emph{sub-samples}. 
We took advantage of the fact
that quasars are observationally tagged with the number of the plate 
on which they were observed.
A given pair belongs to the sub-sample $p$ if the quasar with the 
smaller right ascension in the pair 
was observed at plate $p$. Thus there are as many sub-samples as 
the number of plates ($\mathrm{N_{plates}}$) that 
compose the data sample ($\mathrm{N_{plates}}=2044$ for DR11).

In terms of this partition of the data sample into sub-samples, we
write our estimator of the correlation function in Eq. \ref{eq:thexi} as
\begin{equation}
\hat{\xi}_A ={1\over\sum_{p=1}^\mathrm{N_{plates}}w_A^p} \sum_{p=1}^\mathrm{N_{plates}}\hat{\xi}_A^pw_A^p
\;,
\end{equation}
where $\hat{\xi}_A^p$ is the correlation function calculated using only pairs
belonging to the sub-sample $p$ and $w_A^p$ is the sum of their weights. The
denominator is equal to the sum of weights in $A$, the normalization in
Eq. \ref{eq:thexi}.

Our partitioning scheme ensures that a pair of quasars 
contributes to one and only one sub-sample. 
This approach implies that the correlation 
between $\xi_A^p$ and $\xi_A^{p'}$ (with $p\ne p'$) is given only by terms of
the form T4, T5, and T6. Below we neglect this small
correlation.

The covariance matrix is then given by
\begin{equation}
\langle\hat{\xi}_A\hat{\xi}_B\rangle-\langle\hat{\xi}_A\rangle\langle\hat{\xi}_B\rangle
=S_{AB}^{-1}
\sum_{p=1}^\mathrm{N_{plates}} w_A^pw_B^p\left(\langle\hat{\xi}_A^p\hat{\xi}_B^p\rangle-\langle\hat{\xi}_A\rangle\langle\hat{\xi}_B\rangle\right)
\;,
\end{equation}
where, as anticipated, we assumed that crossed terms from different plates are zero.
The final step is to use the following estimator for the expression in parentheses, the covariance within a plate:
\begin{equation}
\hat{C}_{AB}^p = \hat{\xi}_A^p\hat{\xi}^p_B-\hat{\xi}_A\hat{\xi}_B
\;.
\end{equation}

\input{CrossCovariance}

\section{Correlation function for C1, C2, and C3}
\label{wedgeappendix}

\begin{figure*}[tb] 
\includegraphics[width=\textwidth]{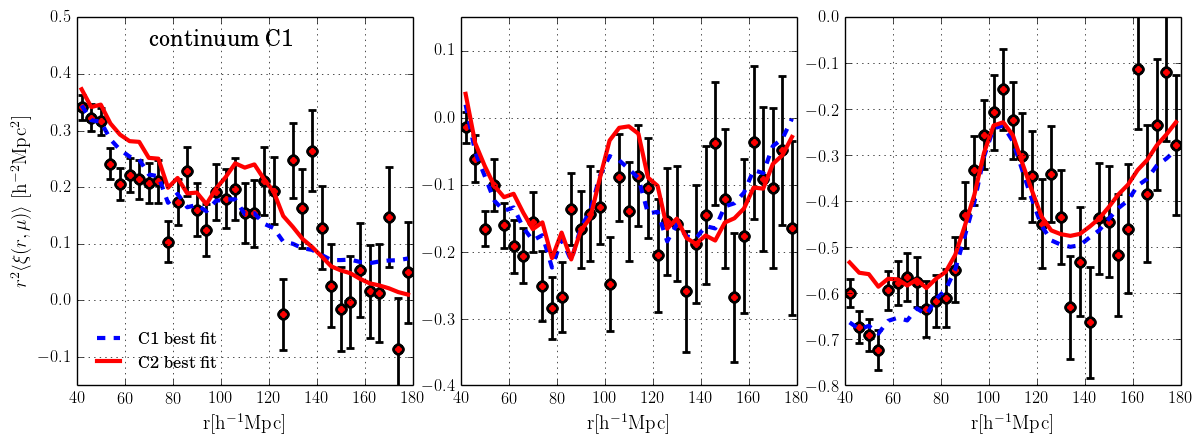}
\includegraphics[width=\textwidth]{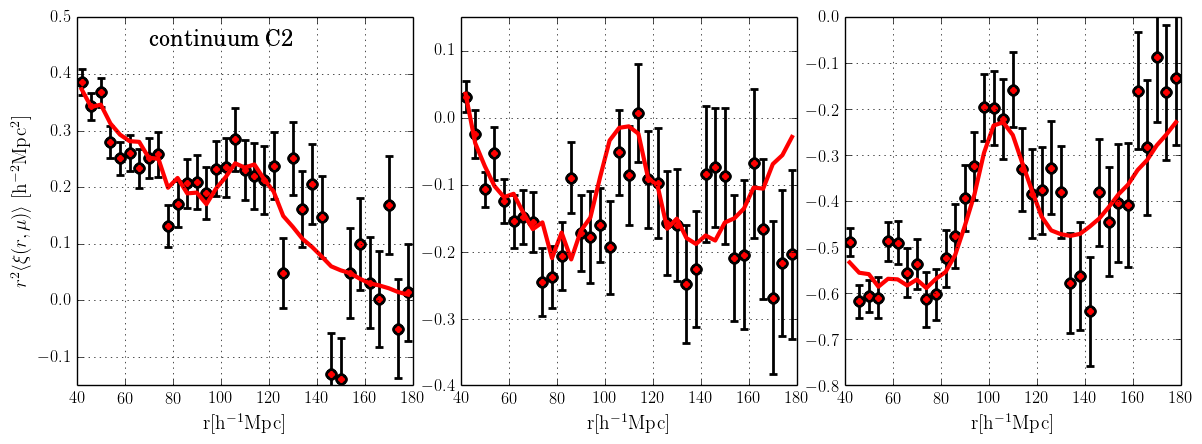}
\includegraphics[width=\textwidth]{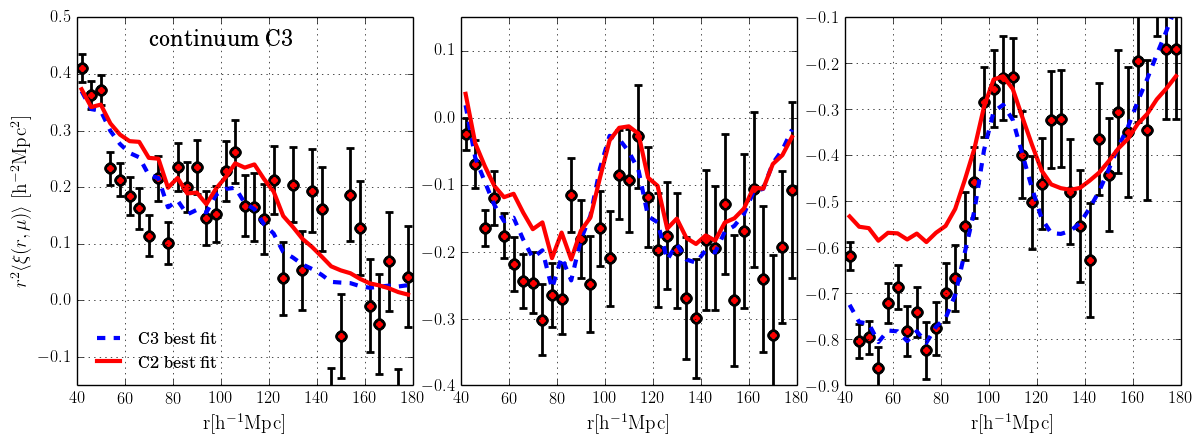}
\caption{
Measured correlation functions for continuum C1 (top),
C2 (middle), and C3 (bottom) averaged over
three angular regions: 
$0<\mu<0.5$ (left),
$0.5<\mu<0.8$ (middle), and 
$\mu>0.8$ (right),  
The curves show the results of fits as described in Sect. \ref{fitssec}.
The blue dashed curves are the best fits, the full red curves are the
best fit for C2. 
}
\label{3Cfig}
\end{figure*}

Figure \ref{3Cfig} shows the correlation function found
using the three continuum estimators in three ranges of $\mu$
(the same as those in Fig. \ref{wedgesfig}).
All three methods show a clear BAO peak in the nearly radial bin,
$\mu>0.8$.  While the peak in this bin is at the same position
for all continua, the absolute value of the correlation function is quite 
different.  This is because the $\mu>0.8$ bin is strongly affected
by the distortions induced by the continuum estimator.

\end{document}

%% file: CrossCovariance.tex
\section{Combining the results with those of \citet{font13}}
\label{crosscovariance}

In this appendix we discuss the level of correlation between the BAO 
measurement presented in this paper and that measured in \citet{font13} 
from the cross-correlation of the \Lya forest with the quasar density field,
also using the DR11 of BOSS. 

If both analyses were limited by cosmic variance, there would be no gain in 
combining them, since both would be tracing the same underlying density 
fluctuations. However, as shown in appendix B of \citet{font13}, cosmic variance
is only a minor contribution to the uncertainties in both measurements. The
accuracy of the \Lya autocorrelation measurement (presented here) is limited
by the aliasing noise \citep{mcdoeisen07,mcquinnwhite11} 
and the instrumental noise, while the cross-correlation 
measurement \citep{font13} is also limited by 
the shot-noise of the quasar field. Since the dominant sources of
fluctuation in the two measurements have a completely different nature,
the cross covariance should be small.

To better quantify this statement, we calculate the covariance between the two measurements by
computing the cross-correlation coefficient between a bin measured in the 
autocorrelation $\hat\xi_A$ and a bin measured in the 
cross-correlation $\hat\xi_a$, defined as
\begin{equation}
 r_{Aa} = \frac{C_{Aa}}{\sqrt{C_{AA}~C_{aa}}} ~,
\end{equation}
where $C_{AA}$ is the variance in the autocorrelation bin $A$, $C_{aa}$ is 
the variance in the cross-correlation bin $a$, and $C_{Aa}$ is the covariance
between the two bins. 

\begin{figure}[htb]
\includegraphics[width=.47\textwidth]{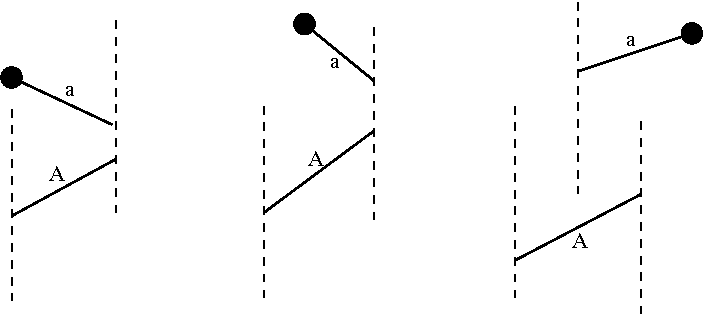}
\caption{Three types of diagrams constributing to the covariance between a 
bin $A$ in the \Lya autocorrelation and a bin $a$ in the cross-correlation with
quasars. The dashed lines refer to \Lya forests, the dots to quasars.
The solid lines refer to \Lya pixel pairs or quasar-\Lya pairs used
to measure the auto- or cross-correlation.}
\label{cross_graphs}
\end{figure}

We calculated the covariance $C_{Aa}$ using a Wick expansion similar to that
computed in appendix \ref{covappendix}. 
In this case, we must compute a four-point
function with three \Lya pixels and a quasar position, and the different
contributions to the covariance will be products of the \Lya autocorrelation
function between two pixels and the \Lya-quasar cross-correalation between a
quasar and a pixel. As shown in Fig. \ref{cross_graphs}, there will be 
three types of contribution to the covariance, arising from configurations 
with two, three, and four quasars.

The correlation of pixels in different lines of sight will in general be
weaker than the correlation of pixels in the same line of sight. Therefore, 
we expect the right-most diagram in Fig. \ref{cross_graphs} to have a small
contribution, since it involves pixel pairs from different lines of sight.

Direct compuation shows that the contribution from three quasar diagrams is about 
a factor of ten higher than that from two-quasar diagrams for $r^A_\perp=r^a_\perp$.
The two-quasar contribution for $r^A_\perp\neq r^a_\perp$ is zero.

\begin{figure}[htb]
\includegraphics[width=.47\textwidth]{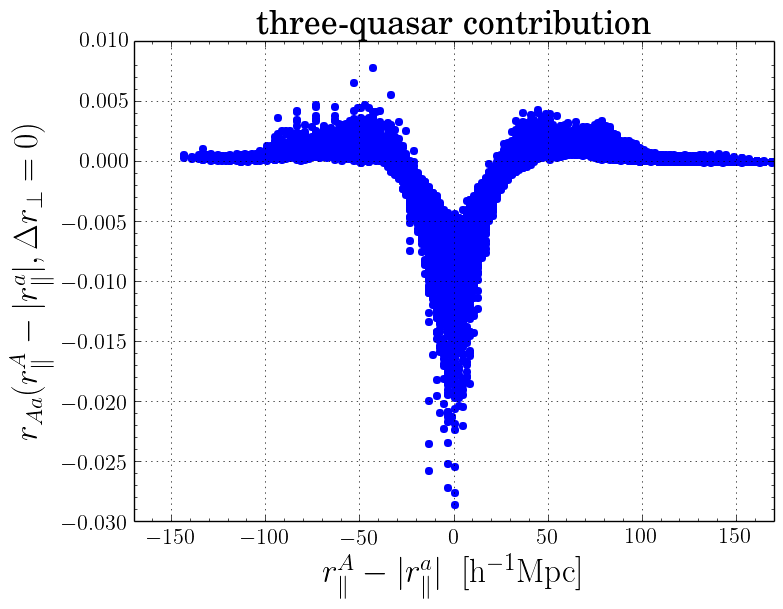}
\caption{Cross-correlation coefficients as a function of $\rpar^A-|\rpar^a|$,
computed from the three-quasar diagrams.
$A$ and $a$ refer to bins for the autocorrelation
and cross-correlation measurements.}
\label{cross_covar}
\end{figure}

In Fig. \ref{cross_covar} we show the cross-correlation coefficients computed
from the three-quasar diagrams as a function of $\rpar^A-|\rpar^a|$ for $\rperp^A=\rperp^a$ 
(it decreases with increasing $|\rperp^A-\rperp^a|$). As expected, 
the correlation between the two measurements is weak, justifying the combined 
contours presented in Fig.
\ref{crossautofig}.